\documentclass{article}

\usepackage[english]{babel}

\usepackage[letterpaper,top=2cm,bottom=2cm,left=3cm,right=3cm,marginparwidth=1.75cm]{geometry}

\usepackage{amsmath}
\usepackage{graphicx}
\usepackage{natbib} 
\usepackage[colorlinks=true, allcolors=blue]{hyperref}

\title{Vertical Distribution of Optical Turbulence at the Peak Terskol Observatory and Mount Kurapdag}
\author{Artem Yu. Shikhovtsev $^{1,*}$ , Chun Qing $^{2,3,*}$ , Evgeniy A. Kopylov $^{4}$, Sergey A. Potanin$^{5,6}$, 
\\ Pavel G. Kovadlo$^{1}$}

\begin{document}
\maketitle

\begin{abstract}
Characterization of atmospheric turbulence is essential to understanding image quality of astronomical telescopes and applying adaptive optics systems. In this study, the vertical distributions of optical turbulence at the Peak Terskol Observatory (43.27472 ° N 42.50083 ° E, 3127 m a.s.l.) using the Era-5 re-analysis, scintillation measurements and sonic anemometer data are investigated.  For the reanalysis grid node closest to the observatory, vertical profiles of the structural constant of the air refractive index turbulent fluctuations $C_n^2$ were obtained. The calculated $C_n^2(z)$ vertical profiles are compared with the vertical distribution of turbulence intensity obtained from tomographic measurements with Shack-Hartmann sensor. The Fried parameter $r_0$ at the location of Terskol Peak Observatory was estimated. Using combination of atmospheric models and scheme paramaterization of turbulence, $C_n^2(z)$ profiles at Mt. Kurapdag were obtained. The $r_0$ values at the Peak Terskol Observatory are compared with estimated values of this length at the ten astronomical sites including Ali, Lenghu and Daocheng.
\end{abstract}

\section{Introduction}

Adaptive optics systems correcting optical distortions in a wide field of view require information about vertical profiles of optical turbulence and wind speed. The key characteristics of the optical turbulence are the structural constant of the air refractive index turbulent fluctuations $C_n^2$ and the atmospheric coherence length, often associated with the Fried parameter $r_0$ and seeing \citep{Coulman1985, Tokovinin2023}. These quantities determine the possibilities of reaching a high angular resolution in astronomical observations. 

There are a number of techniques for measurements of optical turbulence vertical profiles. In astronomical observations, the vertical profiles of optical turbulence above telescope can be determined with a multidirectional Shack–Hartmann wavefront sensor \citep{Ran2024}. In particular, the vertical profiles of $C_n^2$ are calculated by applying the Solar Differential Image Motion Monitor (S-DIMM+) or Slope Detection and Ranging (SLODAR) techniques \citep{Hickson2019, Hickson2020, Wang2018, Subramanian2023, Shikhovtsev2022,Joo2024,Bennoui2023, Deng2023, Wilson}.  The most complete review of methods for measuring optical turbulence vertical profiles is given in a recent paper \citep{Griffiths2024}. Griffiths R. et al. present the results of a intercomparison of turbulence profiling instruments including Stereo Scintillation  Detection and Ranging (S-SCIDAR), DIMM, 24-hour Shack-Hartmann Image Motion Monitor (24hSHIMM), Ring-Image Next Generation Scintillation Sensor (RINGSS), Multi-Aperture Scintillation Sensor-Differential Image Motion Monitor (MASS-DIMM). It is important to emphasize that, the comparison of these turbulence profiling instruments showed that the largest deviations correspond to estimates of the isoplanatic angle $\theta_0$, the values of which are highly sensitive to turbulence in the upper atmospheric layers. These deviations depicted in the form of scattered points are due to a limited accuracy in determination of the optical turbulence strength at different atmospheric levels and, apparently, differences in radiation propagation paths in measurements. 

Measurement of optical turbulence vertical profiles is a hard procedure \citep{Zhang2022}. Most often, the profile measurements are carried out only for a separate set of nights or days. In order to estimate the optical turbulence characteristics over long time intervals or, in other words, for a statistical ensemble, special atmospheric models are widely used \citep{Bol2019,Qing2020, Bi2023, Yang2021, Quatresooz2023, Wu2024}. The atmospheric models are applied to estimate the characteristics of mesoscale, micrometeorological and optical turbulence, on spatial scales comparable to the aperture diameters of ground-based telescopes. Below, a few recent papers demonstrating interesting results in this direction are discussed. A very interesting approach to predict optical turbulence was demonstrated \citep{Cuevas2024}. Cuevas et al applied a combination of models to improve the forecasts of $C_n^2$ profiles and seeing at Paranal using Weather Research and Forecasting model. Also, Macatangay et al. \citep{Macatangay2024} utilized advanced numerical simulations with the Weather Research and Forecasting model to predict astronomical seeing. Meteorological model used by the authors reproduced variations very well, especially in comparison with smoothed DIMM data.
Using the 21-yr European Centre for Medium-Range Weather Forecasts’ fifth set of reanalysis (Era-5) data, the optical turbulence and the wind speed characteristics were simulated at the Lenghu site \citep{Bi2024}. Analysis of the optical turbulence statistics shows that calculated median value of seeing is slightly lower than measured seeing. In particular, Era-5 derived value is 0.72 $''$ and measured seeing is 0.75 $''$. Also, in our opinion, significant fundamental results in the optical turbulence diagnostics and forecasting with atmospheric models were obtained by English scientific teams: new methods and tools were created for studying optical turbulence \citep{Masciadri2001A,Turchi2017}. 

For estimating optical turbulence characteristics with atmospheric models, a key step is the selection of a small-scale turbulence parameterization scheme \citep{Shikhovtsev2024PASJ}. Correct selection and adjustment of the parameterization scheme to the atmospheric conditions of a given site are required to obtain representative estimates of the key characteristics of optical turbulence. Within mountainous regions, including Tibet and the Caucasus, turbulence behavior is complex. The choice of parameterization scheme for mountain regions should be based on analysis of preliminary measurement data. We should emphasize also that modern science still needs to study dependencies in turbulence evolution and determine to what extent these dependencies corresponding to a given mountainous region can be applied for another region. 

\begin{figure}
\includegraphics[scale=0.95]{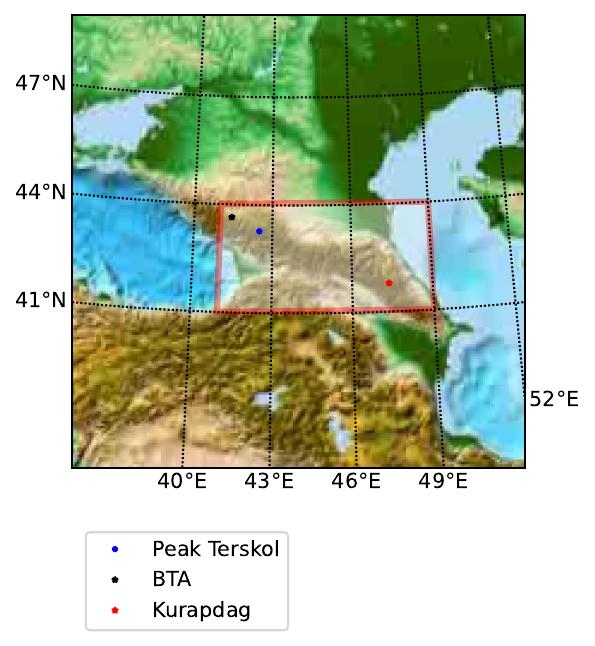}
\caption{Sites of the Terskol Peak Observatory, Big Telescope Alt-Azimuthal Telescope (BTA), Gunib with high astroclimatic indicators and Mount Kurapdag with low water vapor content. Red rectangle shows the BTA region}
 \label{Turbulfgencfggeg}
\end{figure}

The present study focuses on the simulation of optical turbulence above the Peak Terskol Observatory (43.27472 ° N 42.50083 ° E, 3127 m a.s.l.) and Mt. Kurapdag (41.79609 ° N, 47.37428 ° E, 3553 m a.s.l.). These sites were chosen as reference locations, for which data of special atmospheric measurements are available \citep{Bolbasova2023}. The sites of the Terskol Peak Observatory with high astroclimatic characteristics, Mount Kurapdag with low water vapor content \citep{KhaikinESMT} as well as the Big Telescope Alt-Azimuthal Telescope (BTA), the largest optical telescope with a non-segmented mirror in Eurasia are shown in figure \ref{Turbulfgencfggeg}. \citep{Shikhovtsev2024}. Recent studies show that new suitable locations for ground based telescopes include mountain regions of Dagestan. In this regard, the figure shows the location of the mountain village of Gunib as a place with potentially high characteristics.

In March 2023,  the special experiments at the Peak Terskol Observatory have been carried  to characterize the optical turbulence. Together with the measurements of optical turbulence profiles and meteorological characteristics in the surface layer,  variations of $C_n^2(z)$ have been simulated using a gradient method. Below, we discuss the results of estimating optical turbulence profiles using vertical shears of wind speed, mast meteorological measurements and measured height distributions of optical turbulence.

\section{Data}

This study aims to estimate the image quality in nighttime at the Terskol Peak Observatory, where optical turbulence monitoring is carried out, and at a new location, on the slope of Mt.Kurapdag, considered as one of the reference sites for methodological tests. In this study, we solved a number of tasks. Firstly, using remote telescopic measurements we determined vertical distributions of the optical turbulence intensity over the Peak Terskol observatory. Vertical distributions of turbulence intensity were obtained under conditions of good and mediocre image quality. 
Secondly, in order to estimate the vertical profiles of the structural constant of air  refractive index turbulent fluctuations $C_n^2$, we used a gradient method, the foundations of which have been created by Tatarsky. Using the gradient method, the vertical profiles of optical turbulence over the Terskol Peak Observatory and Mt. Kurapdag were determined. 

For simulation of $C_n(z)$ vertical profiles, we used the European Centre for Medium-Range Weather Forecasts’ fifth set of reanalysis (Era-5) data \citep{Hersbach2020}. The Era-5 reanalysis contains observational data from around the world from 1940 to the present. Compared to previous databases, Era-5 demonstrates significant improvements in both spatial resolution and accuracy of atmospheric characteristics \citep{Huang2021, Rao2024}. The re-analysis contains assimilated meteorological characteristics including air temperature and wind speed components at various pressure levels. The meteorological characteristics are available with a flat horizontal resolution of about 30 km and a continuous 1-hour temporal resolution. In particular, for the grid node closest to the observatory, hourly values of air temperature and horizontal components of wind speed at various pressure surfaces were used.

\section {Results}
\subsection{Optical turbulence vertical distributions from Shack-Hartmann sensor measurements at the Peak Terskol Observatory}

\begin{figure}
\includegraphics[scale=0.43]{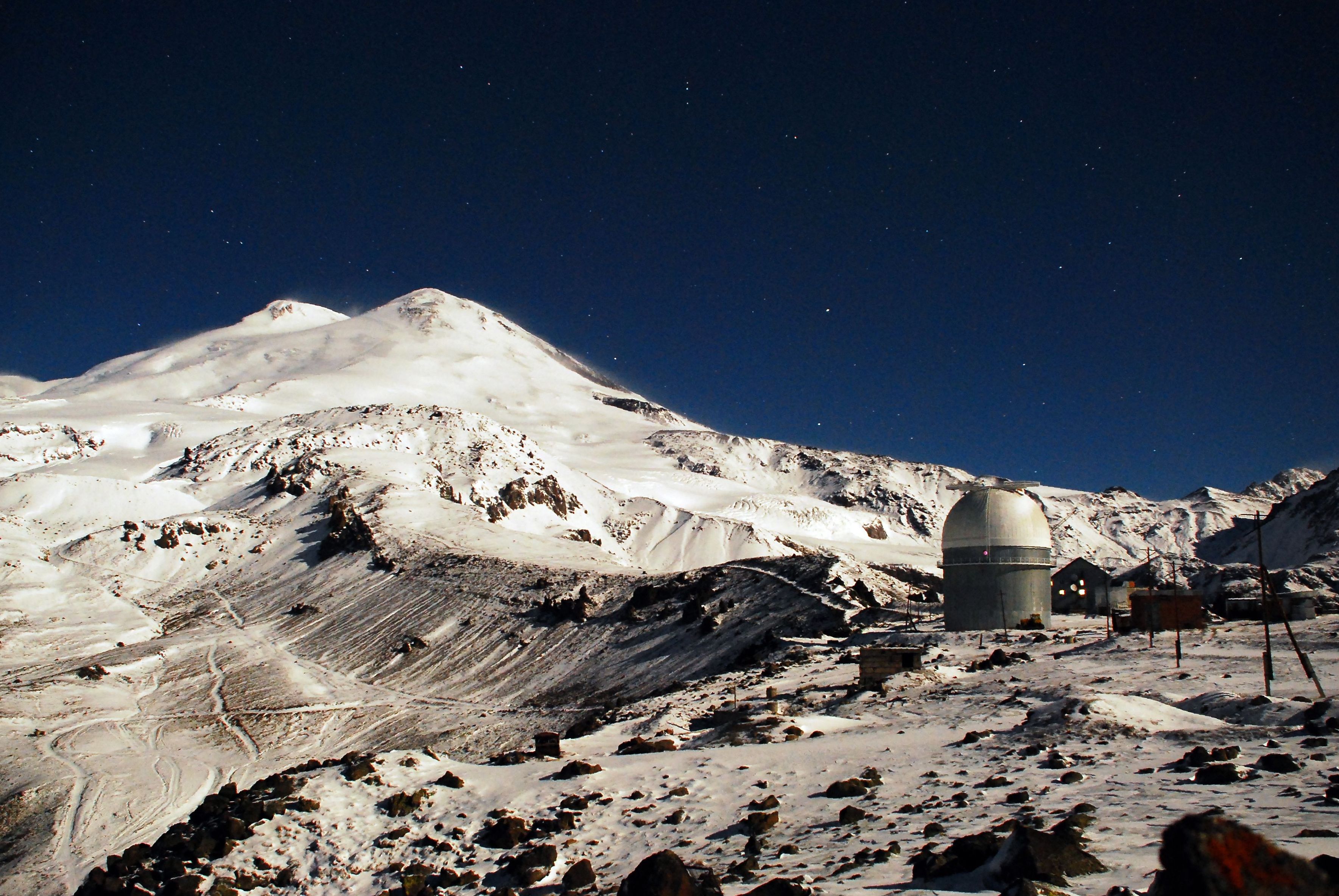}
\caption{The dome of Schmidt-Cassegrain Telescope at the Peak Terskol Observatory}
 \label{Telescope Terskol}
\end{figure}

In solar astronomy,  Shack–Hartmann wavefront sensors with different fields of views are typically employed for estimated wavefront distortions\citep{Townson2015}. The parameters of standard sensor depend on the optical turbulence characteristics. In the paper, we present the results of an observing campaign of the Shack-Hartmann sensor at the Peak Terskol Observatory in 2023 March. We used the Shack-Hartmann sensor placed in the focal plane of MEADE LX200 Telescope installed inside the Schmidt-Cassegrain Telescope (figure \ref{Telescope Terskol}).  The sensor is equipped with a high frame-rate FLIR Grasshopper GS3-U3-28S5M-C camera. Subaperture size of the Shack-Hartmann sensor is 6x6 mm. Exposure time is 5 ms. The sampling is 65 frame/sec.  In experiments, we analyzed 20 full time series and 10000 hartmannograms were obtained during the night of 19 March, 2023.  Analysis of these measurement data made it possible to estimate and analyze the character of optical turbulence above the telescope.


\begin{figure}
\includegraphics[scale=0.65]{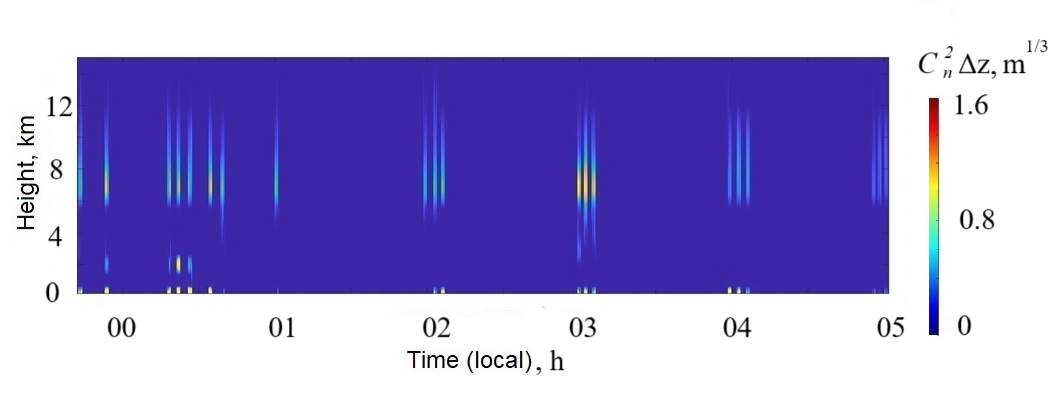}
\caption{Night-time vertical distributions of optical turbulence strength  $C_n^2 \Delta z$ at the Peak Terskol Observatory for the average image quality, 19 March, 2023. In the figure, we used a local time}
 \label{Turbfulfgencfgge}
\end{figure}

\begin{figure}
\centering
\begin{minipage}[h]{0.985\linewidth}
\center{\includegraphics[width=1\linewidth]{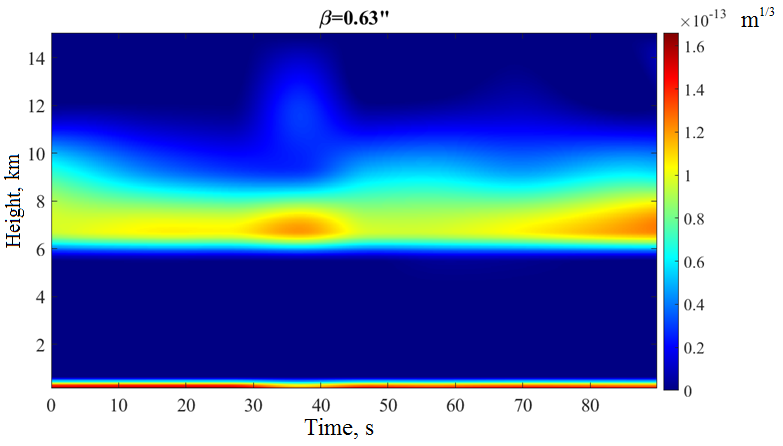}} (\textbf{a})  \ \\
\end{minipage}
\vfill
\begin{minipage}[h]{0.9485\linewidth}
\center{\includegraphics[width=1\linewidth]{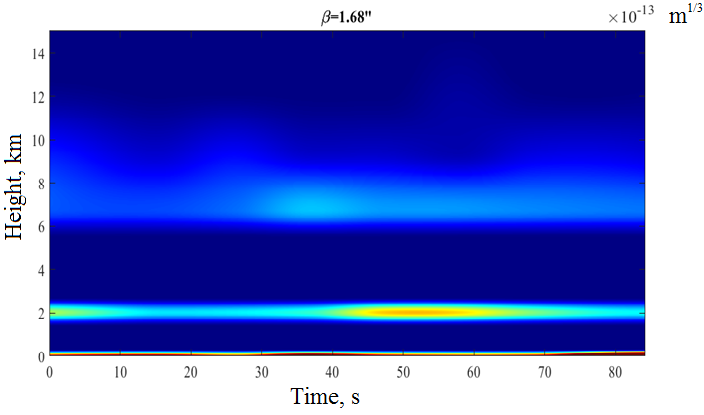}} (\textbf{b})  \\
\end{minipage}
\vfill
\caption{Night-time vertical distributions of optical turbulence strength  $C_n^2 \Delta z$ at the Peak Terskol Observatory for (a)) the best ($\beta$ = 0.63 $''$) and (b)) the worst atmospheric conditions ($\beta$ = 1.68 $''$), 19 March, 2023. In the figure, we used a local time}
\label{Turbulfgence}
\end{figure}

Obtained data sets correspond to average seeing conditions suitable for astronomical observations and thus were selected for processing. Assuming that the turbulence structure obeys the Kolmogorov model, we can estimate the height distribution of optical turbulence strength. For these calculations, we used an expansion of the spatial spectrum of amplitude perturbations into modes. Model spectrum of the scintillation is related to the height distribution of optical turbulence strength: 

\begin{equation}\label{eqcnw}
s_k=\sum_{i=1}^{Nu} W_k(z_i) C_n^2(z_i) \Delta z_i, 
\end{equation}

where $W_k(z_i)$ is the weight function which depends on the turbulent layer height and subaperture geometry, $\Delta z_i$ is the  turbulent layer thickness,  $Nu$ is the number of turbulent layers. Details of the method are described in the paper \citep{Potanin2022}.

y solving this equation in terms of $C_n^2(z_i) \Delta z_i$ we obtain the vertical distributions of optical turbulence strength above the Peak Terskol Observatory. Night-time vertical distributions of optical turbulence strength  $C_n^2 \Delta z$ at the Peak Terskol Observatory for different image quality are shown in figures \ref{Turbfulfgencfgge} - \ref{Turbulfgence}. Image quality is estimated through the  $\beta$ parameter:

\begin{equation}\label{ddddf}
\beta=\frac{0.98 \lambda}{r_0},
\end{equation}

where variance of the differential displacements of the solar subimages $\sigma_{\alpha}^2$ is related to the Fried parameter $r_0$ by the following equation: 
\begin{equation}
\sigma_{\alpha}^2= K_i \lambda^2 r_0^{-5/3}D^{-1/3},
\label{FR}
\end{equation}
where $K_i$ is the numerical constant,  $\lambda$ is the light wavelength and $D$ is the telescope diameter.

Analyzing these figures, we can see that pronounced turbulent layers are formed at heights of 7 - 8 km and 2  - 2.5 km above the ground. The turbulent layer at these heights leads to a significant decrease in image quality. For example, the parameter $\beta$ determined by the integral $C_n^2$ over height increases from 0.63 to 1.68 $''$. These changes are due to, in mainly, generation of intensive turbulent fluctuations of air refractive index within these layers.

It is necessary to emphasize that, conditionally, we consider three different cases, including good seeing ($\beta$ ranges from 0.5 to 1.0 $''$), average seeing ($\beta$ $\sim$ 1.2 $''$) and bad seeing ($\beta$ ranges from 1.5 to 2.0$''$). Good seeing is associated with a thin lower turbulent layer ($\sim$ 50 - 100 m) and the absence of a layer at heights of 2 - 2.5 km. With a thickness of the night boundary layer of 150-250 m, the seeing increases to $\sim$ 1.2 $''$. Values higher than 1.5 $''$ are often associated with the development of optical turbulence within the surface layer and an additional turbulent layer at heights of 2 - 2.5 km.

\subsection{Analysis of sonic measurements within the atmospheric surface layer. Estimation of optical turbulence strength $C_n^2$}

\begin{figure}
\centering
\begin{minipage}[h]{0.655\linewidth}
\center{\includegraphics[width=1\linewidth]{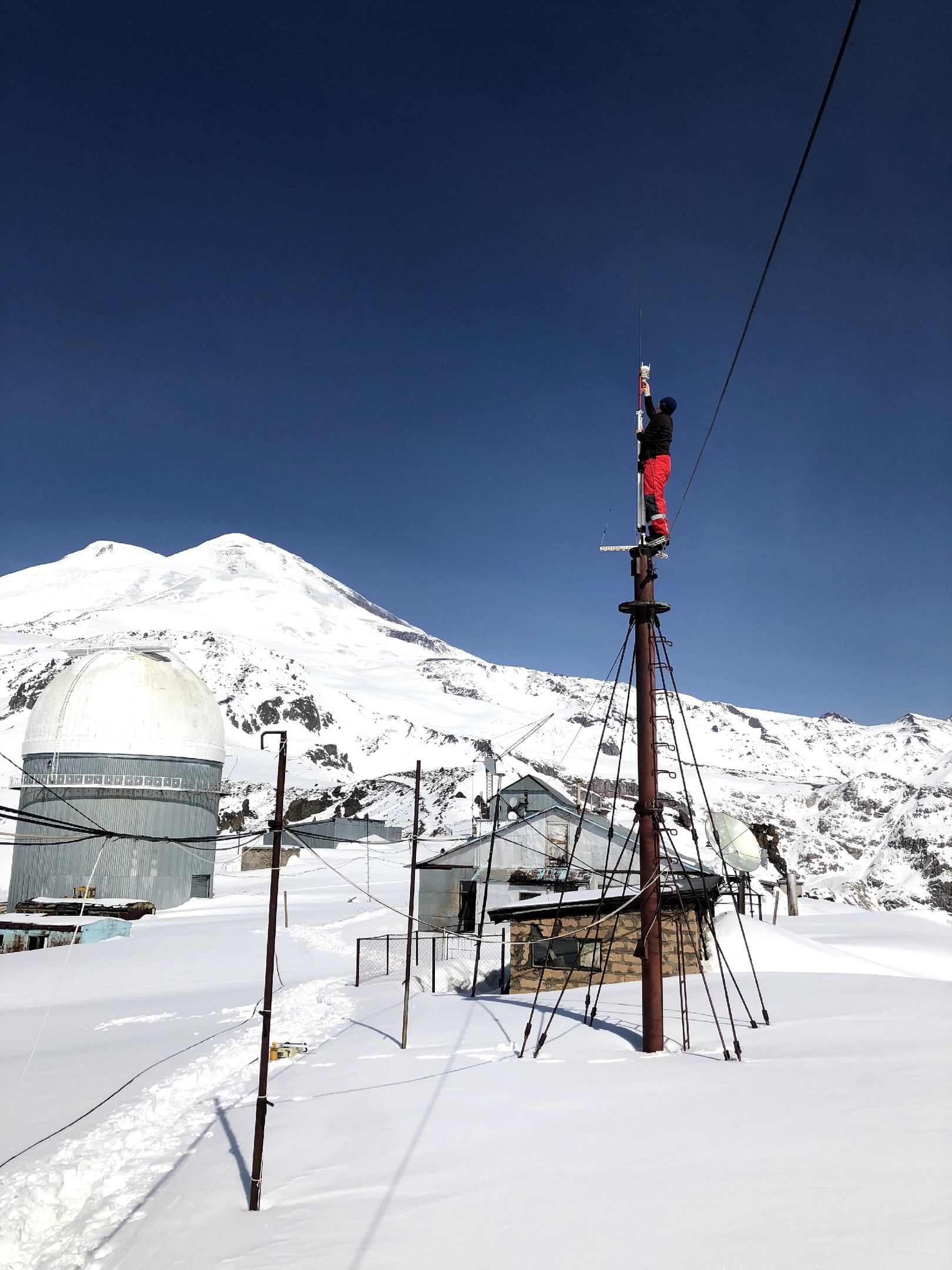}} (\textbf{a})  \ \\
\end{minipage}
\vfill
\begin{minipage}[h]{0.45\linewidth}
\center{\includegraphics[]{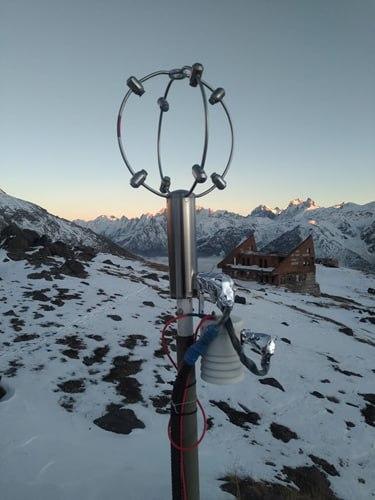}} (\textbf{b})  \\
\end{minipage}
\vfill
\caption{a) The dome of Schmidt-Cassegrain Telescope and b) Meteo-2 sonic anemometer}
\label{efsf}
\end{figure}

The lower part of the atmospheric boundary layer is known makes a major contribution to the formation of integral optical turbulence (along line of sight) and, as a consequence, to the resolving power of ground-based telescope \citep{Kornilov2014,Panchuk2011}. Within this layer, variations in the intensity of optical turbulence are most poorly described,  both in time and in height. At least, the accuracy of optical turbulence estimating and forecasting based on standard atmospheric models or a small amount of optical remote measurement data is not sufficient for planning an observing time. In order to obtain correct shape of optical turbulence profiles within the lowest atmospheric layers under the different atmospheric conditions we analyzed measurement data carried out with a sonic anemometer. The sonic anemometer "Meteo-2" has been placed on a 7-m meteorological mast at the Peak Terskol Observatory site (figure \ref{efsf}). The observatory is located in the vicinity of the Elbrus, on a flat platform at a height of 3100 m above sea level.

Below,  for calculations of atmospheric turbulence characteristics, we used long-period quasi-continuous measurement data of the sonic anemometer (from January 2023 to December 2023). Thanks to the high frequency measurements (10 Hz) of the wind field and the sonic temperature, we calculated a number of the turbulence characteristics including surface values of $C_n^2$.

\begin{figure}
\includegraphics[scale=1.15]{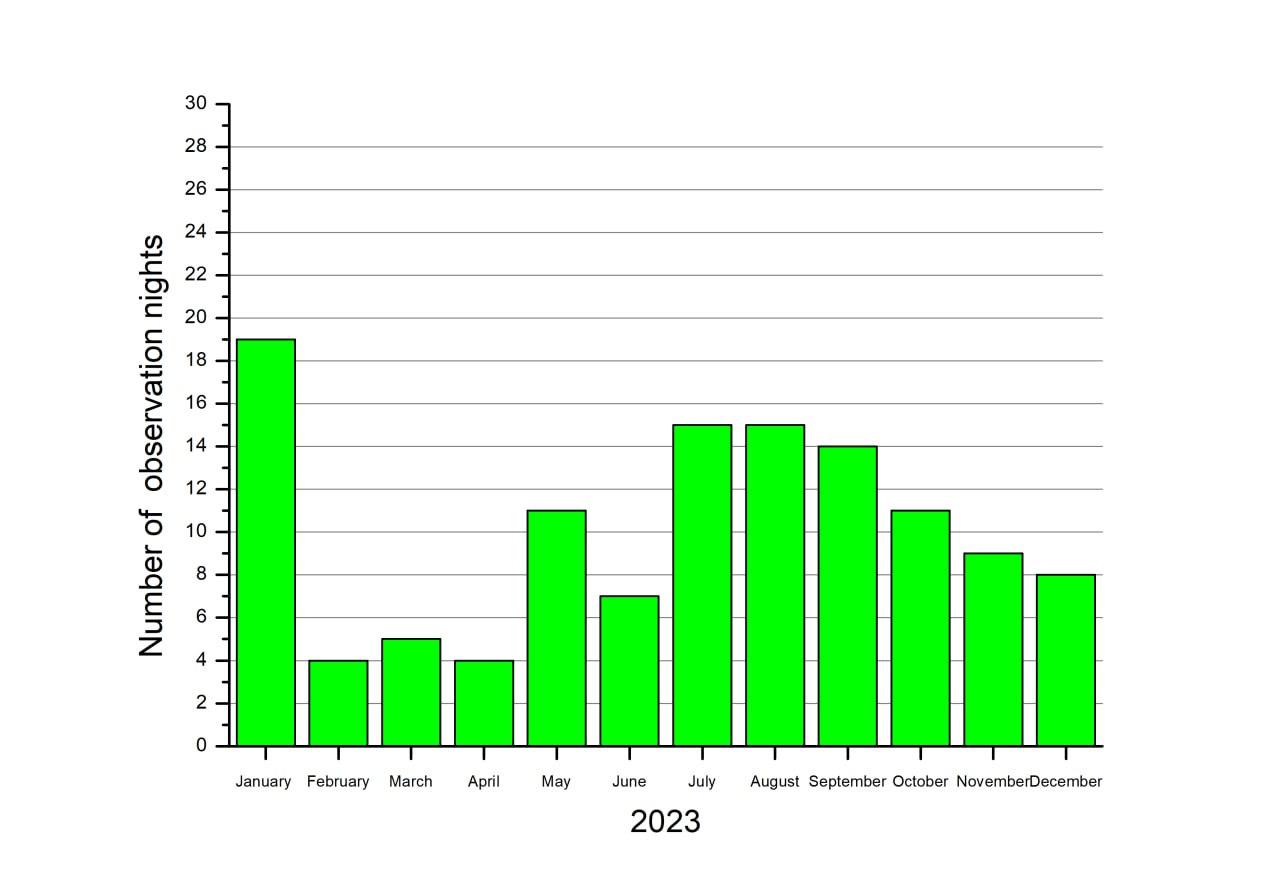}
\caption{The number of nights of sonic measurements by month, 2023}
 \label{Ntr}
\end{figure}

\begin{figure}
\centering
\begin{minipage}[h]{0.529\linewidth}
\center{\includegraphics[width=1\linewidth]{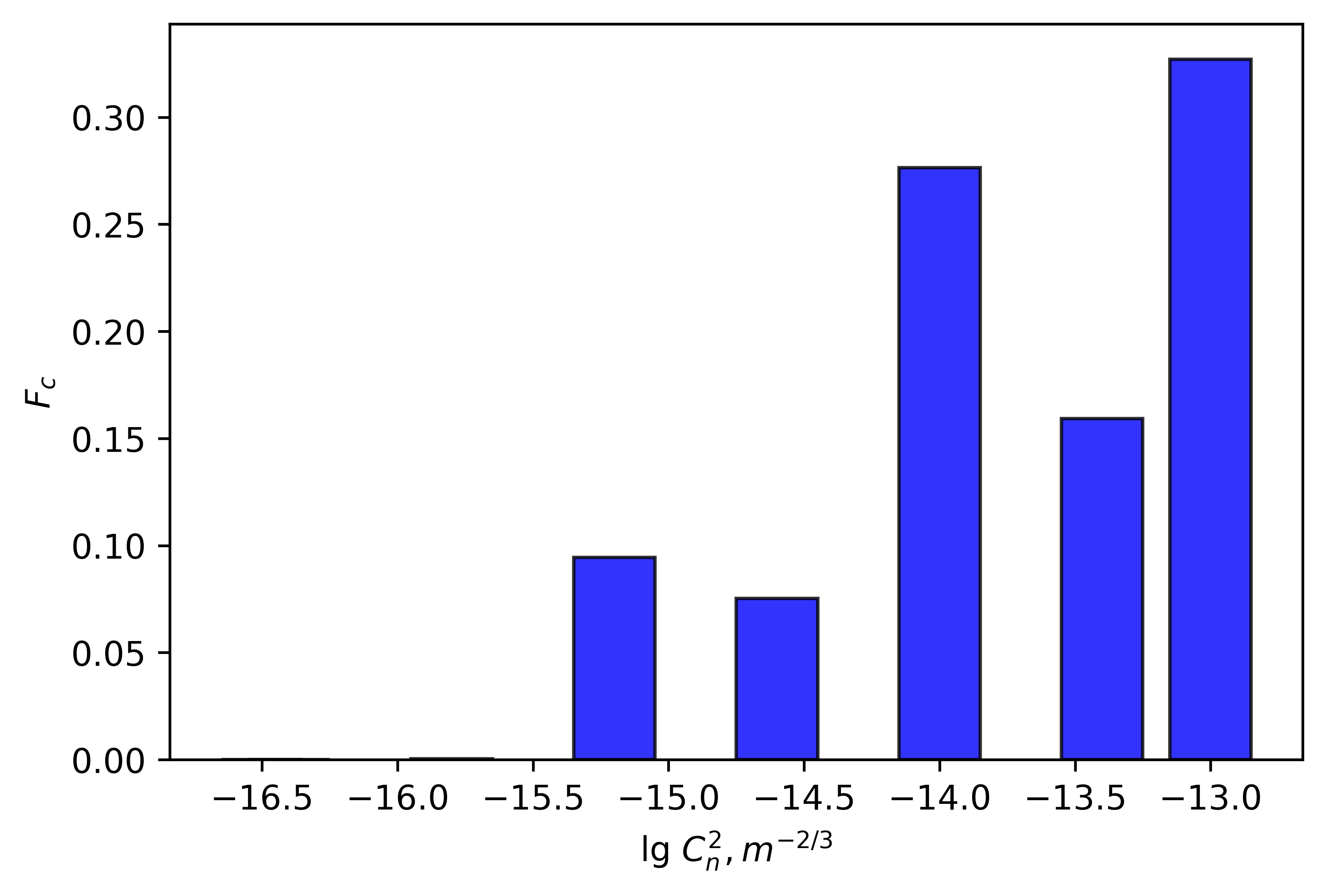}} (\textbf{a}) Winter  \\
\end{minipage}
\hfill
\begin{minipage}[h]{0.529\linewidth}
\center{\includegraphics[width=1\linewidth]{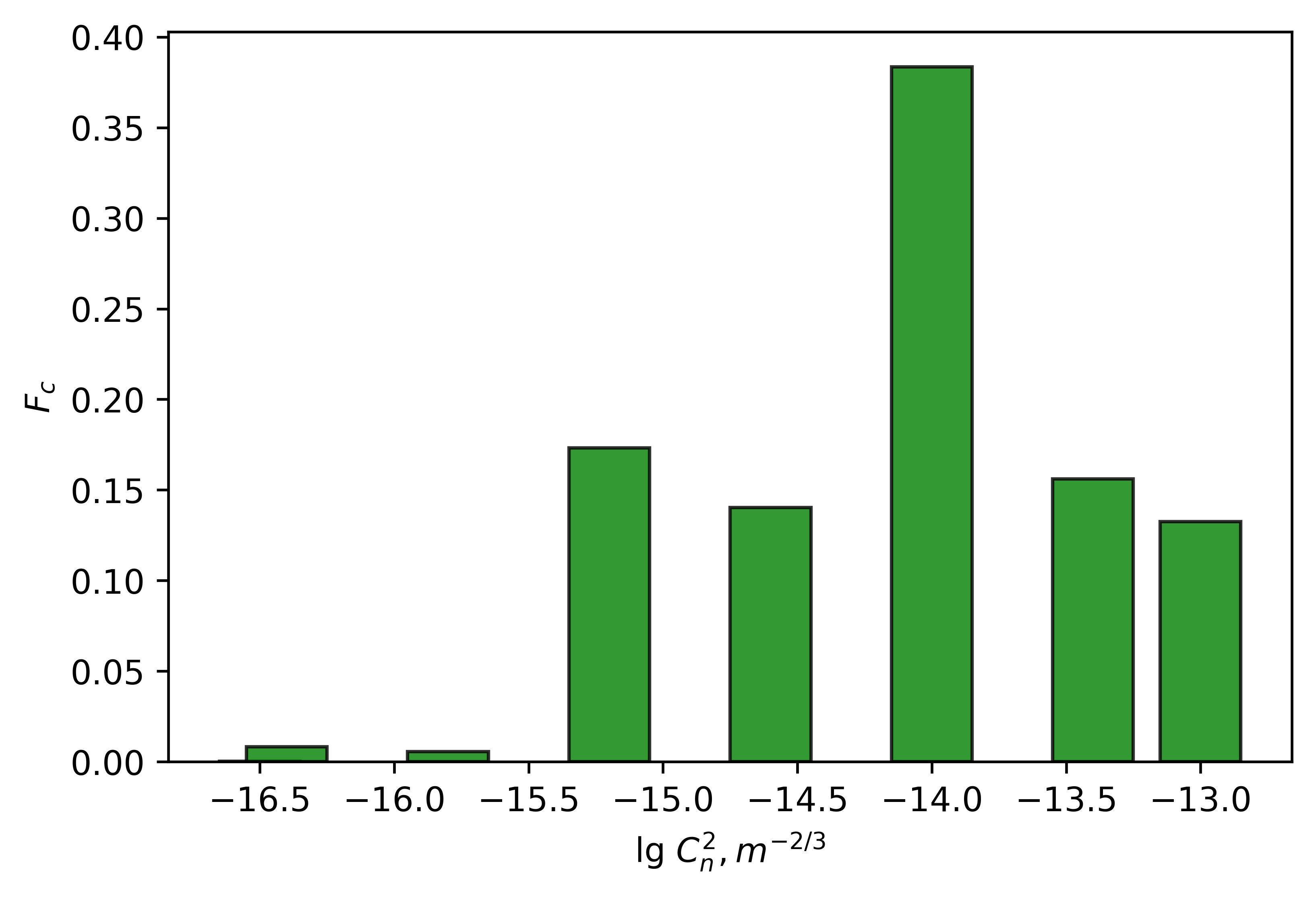}} (\textbf{b}) Spring  \\
\end{minipage}
\hfill
\begin{minipage}[h]{0.529\linewidth}
\center{\includegraphics[width=1\linewidth]{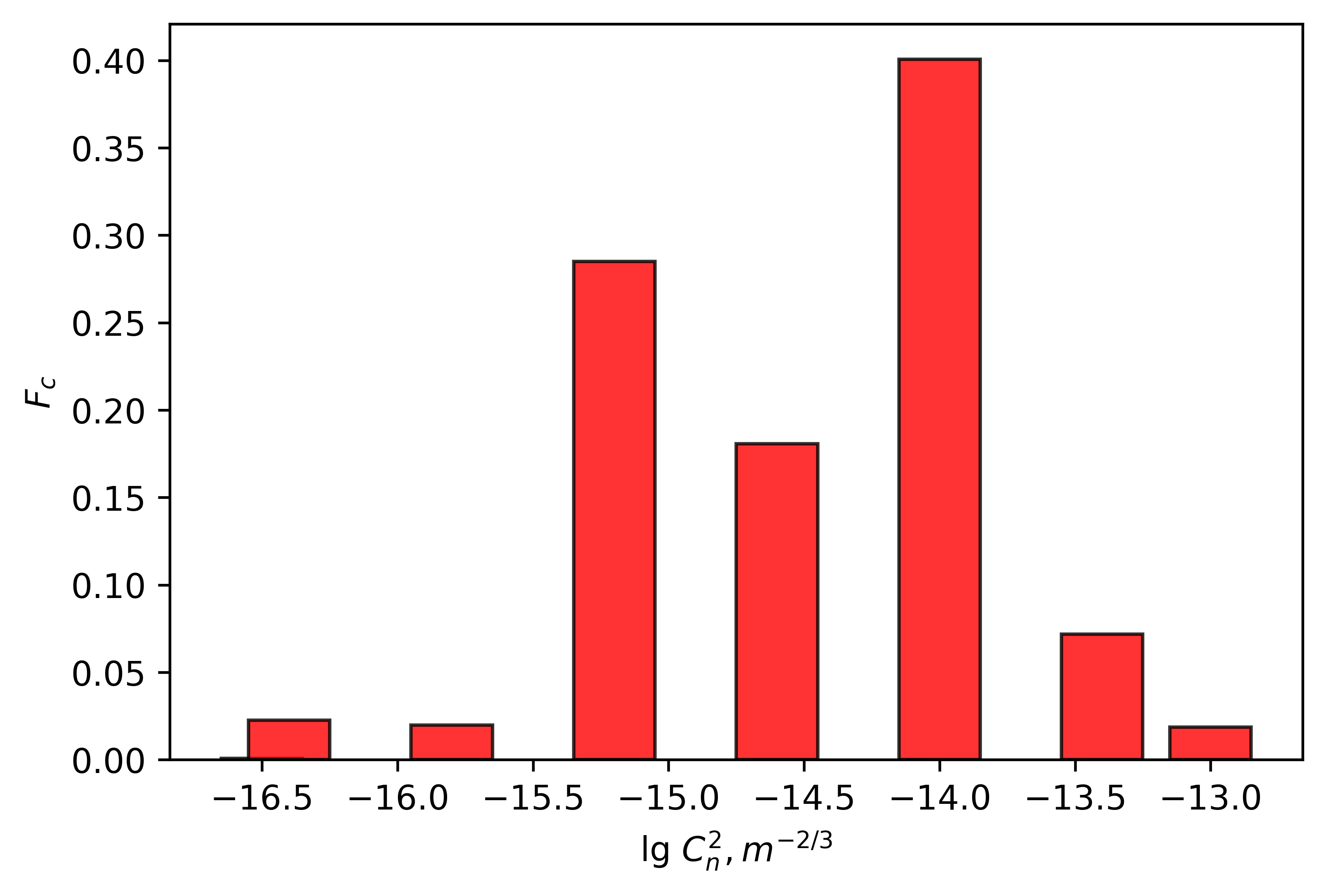}} (\textbf{c}) Summer \\
\end{minipage}
\hfill
\begin{minipage}[h]{0.529\linewidth}
\center{\includegraphics[width=1\linewidth]{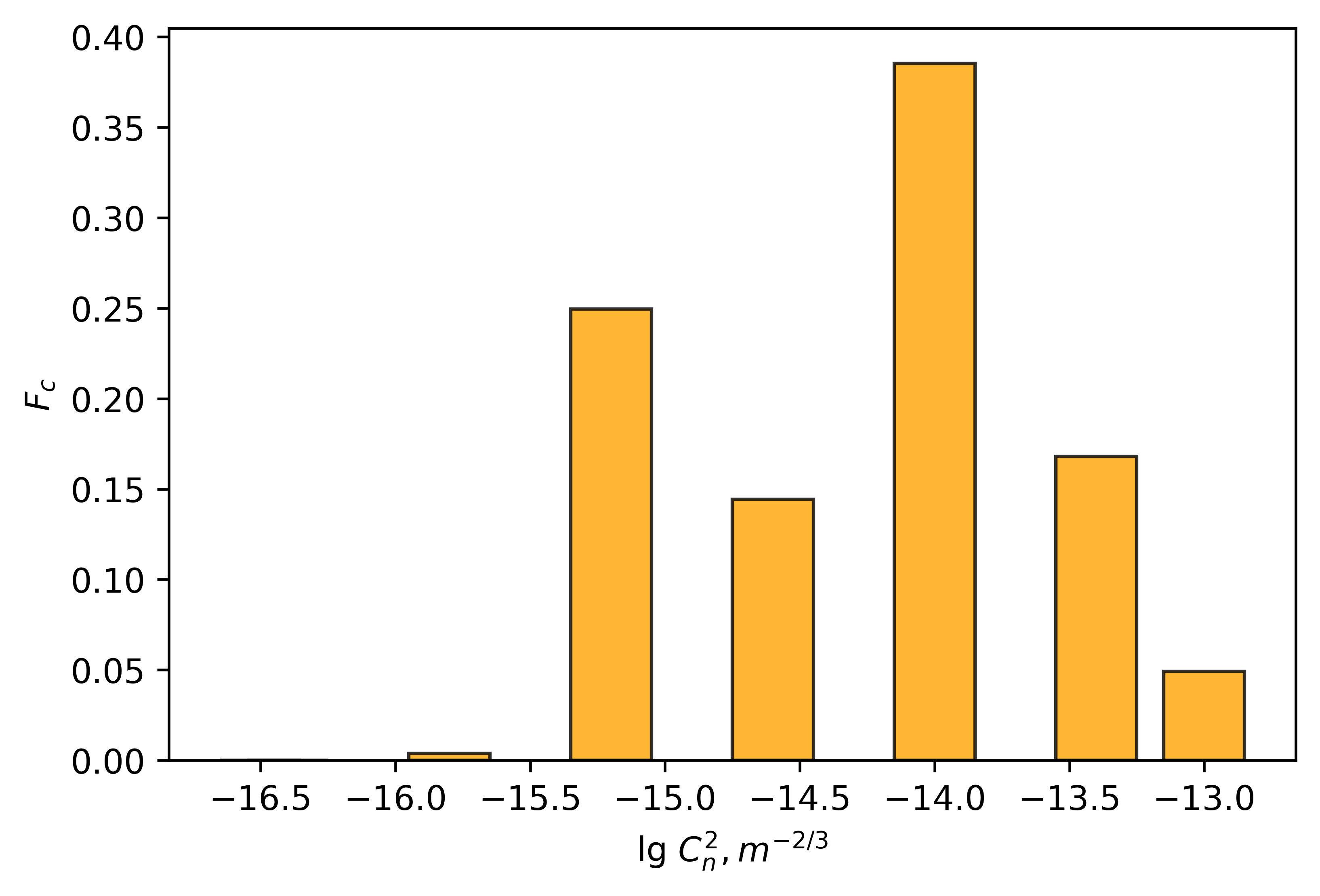}} (\textbf{d}) Autumn  \\
\end{minipage}
\vfill
\caption{Distributions of measured values of $C_n^2$ within the surface layer of the atmosphere at the Peak Terskol Observatory site under average atmospheric conditions, 2023}
\label{2dq}
\end{figure}
Based on the 1-year measurements with sonic anemometer, the 3 minute average values of $C_n^2$ were obtained \citep{Odintsov Atmosphere}. In figure \ref{Ntr}, we present the number of nights sonic measurements for each month. The fewest number of nights analyzed occurs in February-April. The largest number corresponds to January as well as warm period of observations.

\begin{figure}
\centering
\begin{minipage}[h]{0.529\linewidth}
\center{\includegraphics[width=1\linewidth]{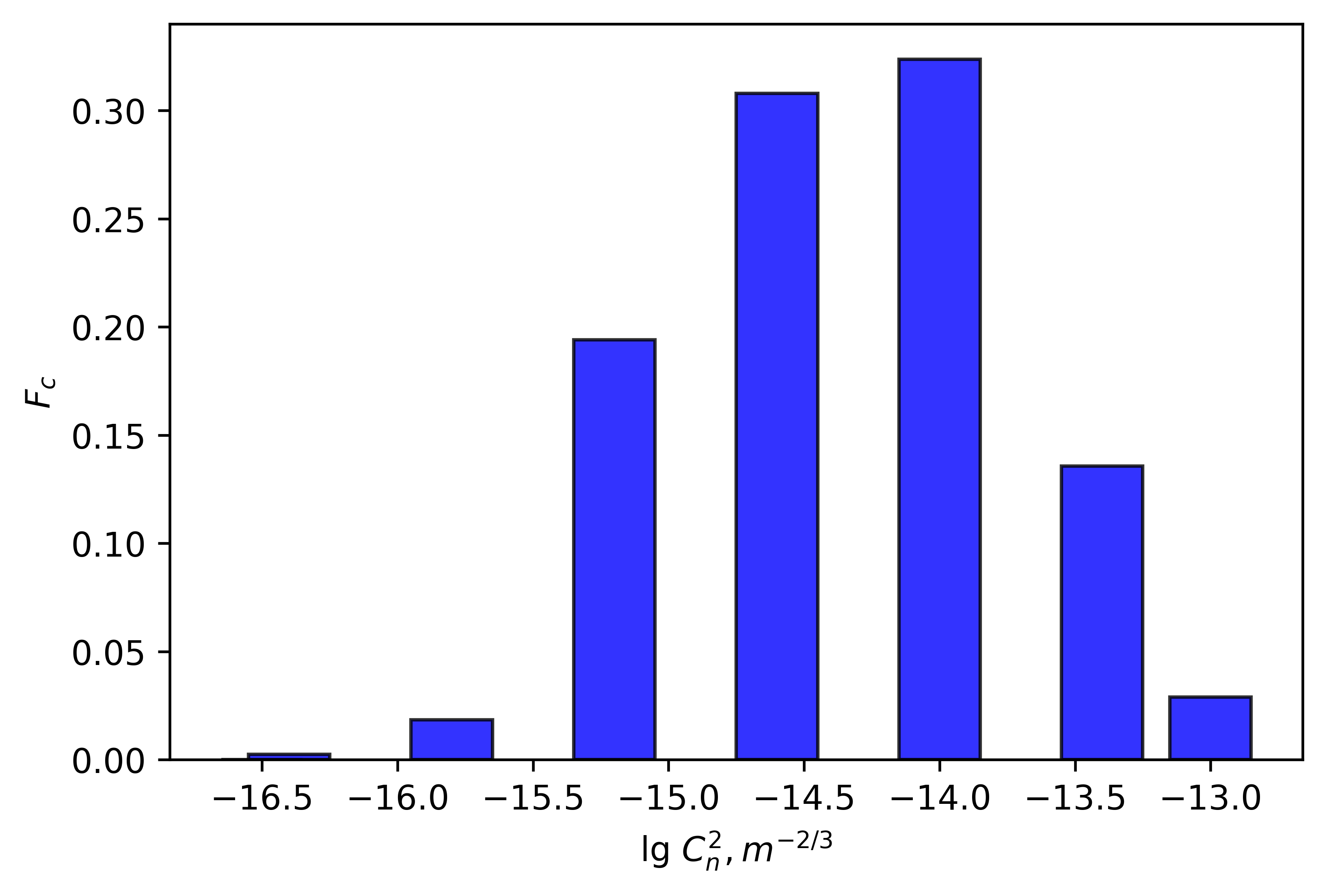}} (\textbf{a}) Winter  \\
\end{minipage}
\hfill
\begin{minipage}[h]{0.529\linewidth}
\center{\includegraphics[width=1\linewidth]{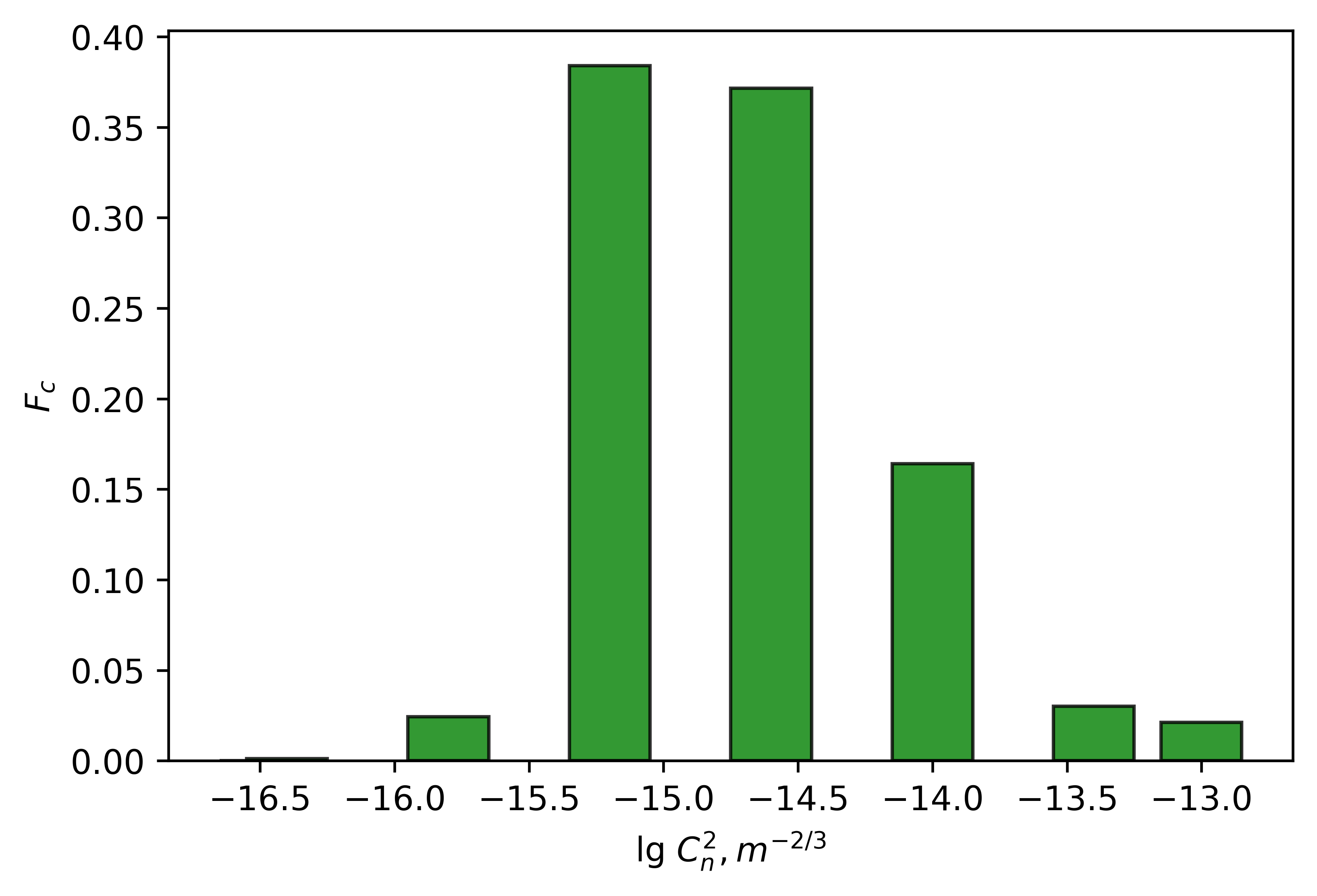}} (\textbf{b}) Spring  \\
\end{minipage}
\hfill
\begin{minipage}[h]{0.529\linewidth}
\center{\includegraphics[width=1\linewidth]{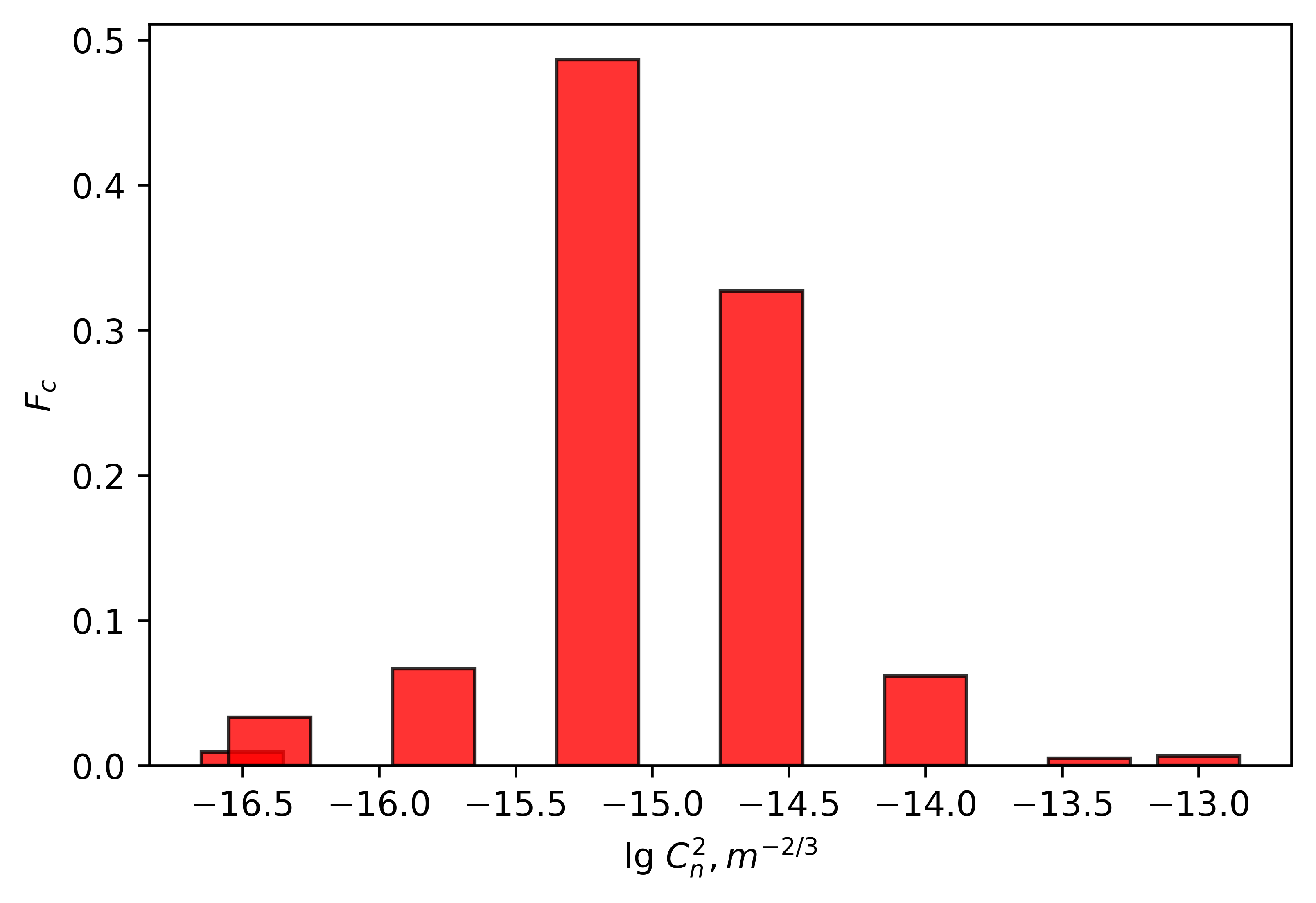}} (\textbf{c}) Summer \\
\end{minipage}
\hfill
\begin{minipage}[h]{0.529\linewidth}
\center{\includegraphics[width=1\linewidth]{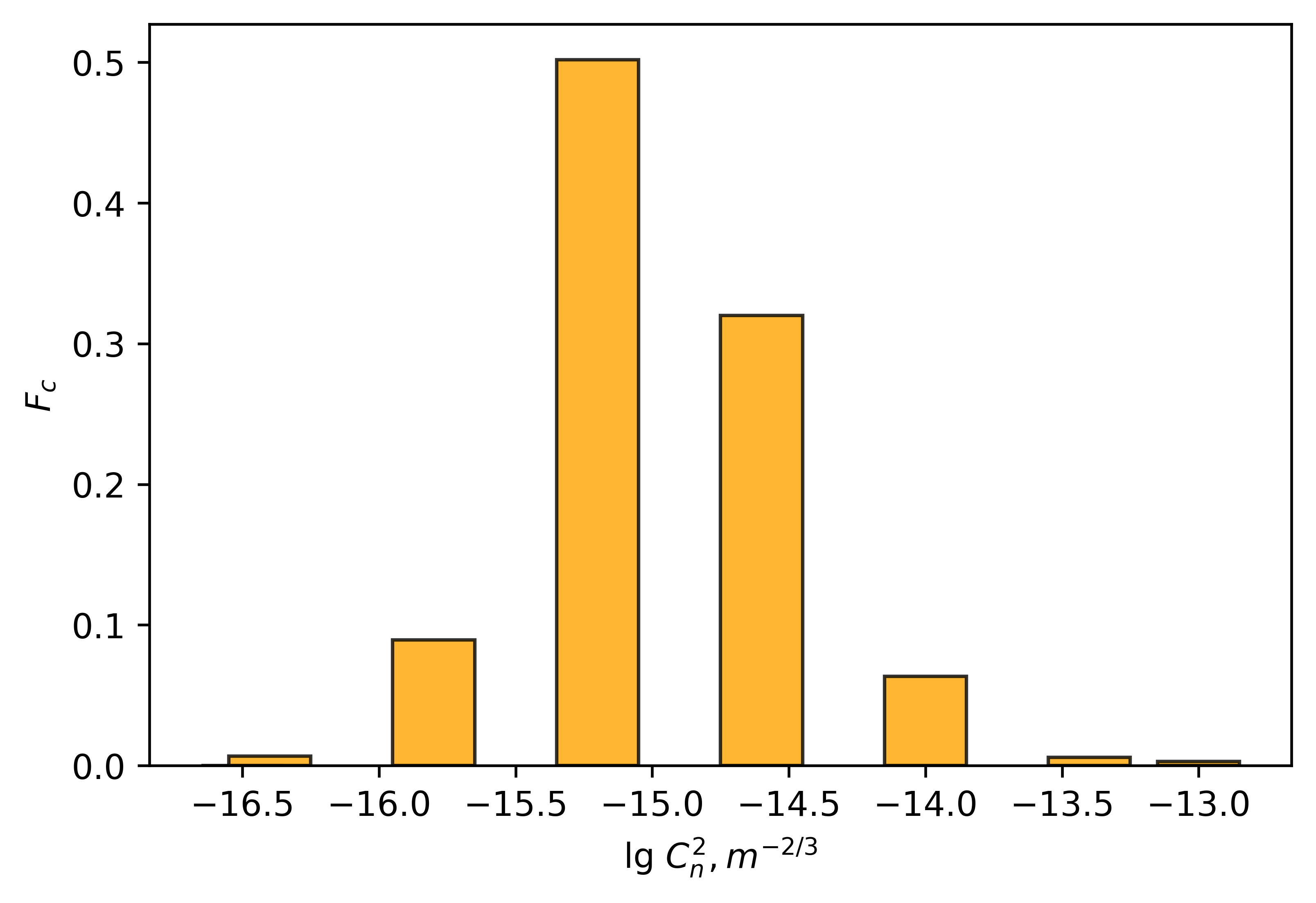}} (\textbf{d}) Autumn  \\
\end{minipage}
\vfill
\caption{Distributions of measured values of $C_n^2$ within the surface layer of the atmosphere at the Peak Terskol Observatory site under clear sky and low intensity of optical turbulence, 2023}
\label{2dg1ccf}
\end{figure}

Distributions of measured values of $C_n^2$ within the surface layer of the atmosphere for different seasons  at the Peak Terskol Observatory site are depicted in figures \ref{2dq} and \ref{2dg1ccf}. 

Table \ref{Ture} contains the calculated percentiles of $C_n^2$ in the surface layer of the atmosphere. 
Calculated values of $C_n^2$ obtained from the results of processing sonic measurement data are higher in comparison with estimations derived from optical measurement data under clear sky conditions. This fact is confirmed by previous studies \citep{Lukin2015}. Higher amplitudes of $C_n^2$ are largely determined by the influence of dynamic processes on the formation of small-scale turbulence in the surface layer of the atmosphere. Due to the fact that astronomical observations are carried out under good astro-optical conditions, when the intensity of optical turbulence along the line of sight is minimal, calculated values in the upper part of table\ref{Ture} differ from statistics corresponding only to clear sky conditions.  In this regard, we evaluated the statistics of surface turbulence for medium atmospheric conditions and low intensity of optical turbulence.

Analysis of estimations shows that the intensity of turbulent fluctuations of the air refractive index has a pronounced seasonal dependence. The highest intensity of optical turbulence in the surface layer of the atmosphere is observed in winter. During this period, the median values of $C_n^2$, estimated for night time (from 0:00 to 6:00 local time), are equal to 8.6 $\cdot$ 10$^{-15}$ m $^{-2/3}$ and 2.6 $\cdot$ 10$^{-15}$ m $^{-2/3}$ for average atmospheric condition and under clear sky,  respectively. In spring, we can an increase in the non-dimensional number of atmospheric situations $F_c$ with low intensity optical turbulence. At this time, during periods of high image quality, the median of $C_n^2$ is about 8.0 $\cdot$ 10$^{-16}$ m $^{-2/3}$. 

In comparison with the winter and spring, in summer and autumn, turbulence in the surface layer is significantly suppressed: the number of atmospheric situations with strong turbulence $C_n^2 \geq $ 10$^{-14}$ m $^{-2/3}$  does not exceed 7\%. We should emphasize that the largest number of cases with the best atmospheric conditions ($C_n^2 \geq $   10$^{-16}$ m $^{-2/3}$ ), occurs in summer and autumn. In this period, about 7 - 10 \% of the observing time corresponds to weak optical turbulence.

\begin{table*}[h]
\caption {Percentiles of $C_n^2$ in the surface layer of the atmosphere at the Peak Terskol Observatory, 2023} \label{Ture}
\medskip
\begin{tabular}{|c|c|c|c|c|}
\hline

Season                        & 10\% , m$^{-2/3}$          & 25\% , m$^{-2/3}$                      & 50\%, m$^{-2/3}$               & 75\%, m$^{-2/3}$                \\
\hline
 \multicolumn{5}{c}{Average atmospheric conditions} \\
\hline
Winter                       & 2.7 $\cdot$  10 $^{-15}$ & 7.0 $\cdot$  10 $^{-15}$           & 8.6 $\cdot$  10 $^{-15}$ & 8.0 $\cdot$  10 $^{-14}$    \\
Spring                         & 4.0 $\cdot$  10 $^{-16}$      & 6.8 $\cdot$  10 $^{-16}$     & 4.4 $\cdot$  10 $^{-15}$     & 1.5 $\cdot$  10 $^{-14}$   \\
Summer                              &  3.0 $\cdot$  10 $^{-16}$           &  5.0 $\cdot$  10 $^{-16}$           & 1.1 $\cdot$  10 $^{-15}$         &  1.2 $\cdot$  10 $^{-14}$        \\
Autumn                             &  3.4 $\cdot$  10 $^{-16}$            &  6.0 $\cdot$  10 $^{-16}$           & 3.8 $\cdot$  10 $^{-15}$         &  3.5 $\cdot$  10 $^{-14}$         \\
 \hline
 \multicolumn{5}{c}{Clear sky, light surface wind $V \leq$ 2.5 m/s } \\
\hline
Winter                       & 5.0 $\cdot$  10 $^{-16}$ & 1.1 $\cdot$  10 $^{-15}$           & 2.6 $\cdot$  10 $^{-15}$ & 1.1 $\cdot$  10 $^{-14}$   \\
Spring                         & 4.8 $\cdot$  10 $^{-16}$      & 5.7 $\cdot$  10 $^{-16}$    & 1.1 $\cdot$  10 $^{-15}$       & 3.1 $\cdot$  10$^{-15}$    \\
Summer                           &  1.4 $\cdot$  10 $^{-16}$           &  4.2 $\cdot$  10 $^{-16}$           & 5.3 $\cdot$  10 $^{-16}$         &  2.2 $\cdot$  10 $^{-15}$         \\
Autumn                         &  1.6 $\cdot$  10 $^{-16}$            &  3.5 $\cdot$  10 $^{-16}$           & 6.0 $\cdot$  10 $^{-16}$         & 3.0 $\cdot$  10 $^{-15}$        \\
\hline
\end{tabular}\\
\end{table*}

\subsection{Vertical distribution of optical turbulence strength estimated from Era-5 reanalysis}

For simulation of the vertical profiles of optical turbulence, the gradient method was used. Correction of the vertical profiles was performed using sonic measurements and optical observations of image scintillation. The calculations were based on the following equation:
\begin{equation}\label{eq1v}
C_n^2(z)=\alpha  L_0(z)^{4/3}M^2 (z),
\end{equation}
where $\alpha$ is the numerical constant which equal to 2.8. The parameter $M(z)$ is the vertical gradient of air refractive index, $L_0$ is the outer scale of turbulence. 

The calculation of the air refractive index gradients $M(z)$  at different heights in the atmosphere is performed using the vertical gradients of the natural logarithm of the air potential temperature $\frac{\partial ln \theta}{\partial z}$:

\begin{equation}
M=\left(\frac{-79 \cdot 10^{-6} P}{T}\right) \frac{\partial ln \theta}{\partial z},
\end{equation}

where $P$ is the atmospheric pressure, $T$ is the air temperature, $\theta$ is the air potential air temperature. The potential air temperature is related to the air temperature by the following formula:
\begin{equation}
\theta = T \left(\frac{1000}{P}\right)^{0.286}.
\end{equation}

The key parameter is the outer scale of atmospheric turbulence $L_0$ \cite{Rao2023}. In general, vertical changes of the outer scale  $L_0$ can be determined through vertical gradients of the horizontal component of the wind velocity $S$:
\begin{equation}\label{eqL0}
L_0^{4/3}= \begin{cases}
    0.1^{4/3} \cdot 10^{a+b \cdot S}, &\text{troposphere} \\
   0.1^{4/3} \cdot 10^{c+d \cdot S}, &\text{stratosphere} 
 \end{cases},
\end{equation}
where
\begin{equation}\label{eqS}
S=\left[\left(\frac{\partial u }{\partial z}\right)^2+\left(\frac{\partial v }{\partial z}\right)^2\right]^{0.5},
\end{equation}

$u$ and $v$ are the horizontal components of wind velocity. Under conditions of weak turbulence the coefficient $a$=1.64, $b$=42, $c$=0.506 and $d$=50. 

\begin{figure}
\centering
\begin{minipage}[h]{0.6185\linewidth}
\center{\includegraphics[width=1\linewidth]{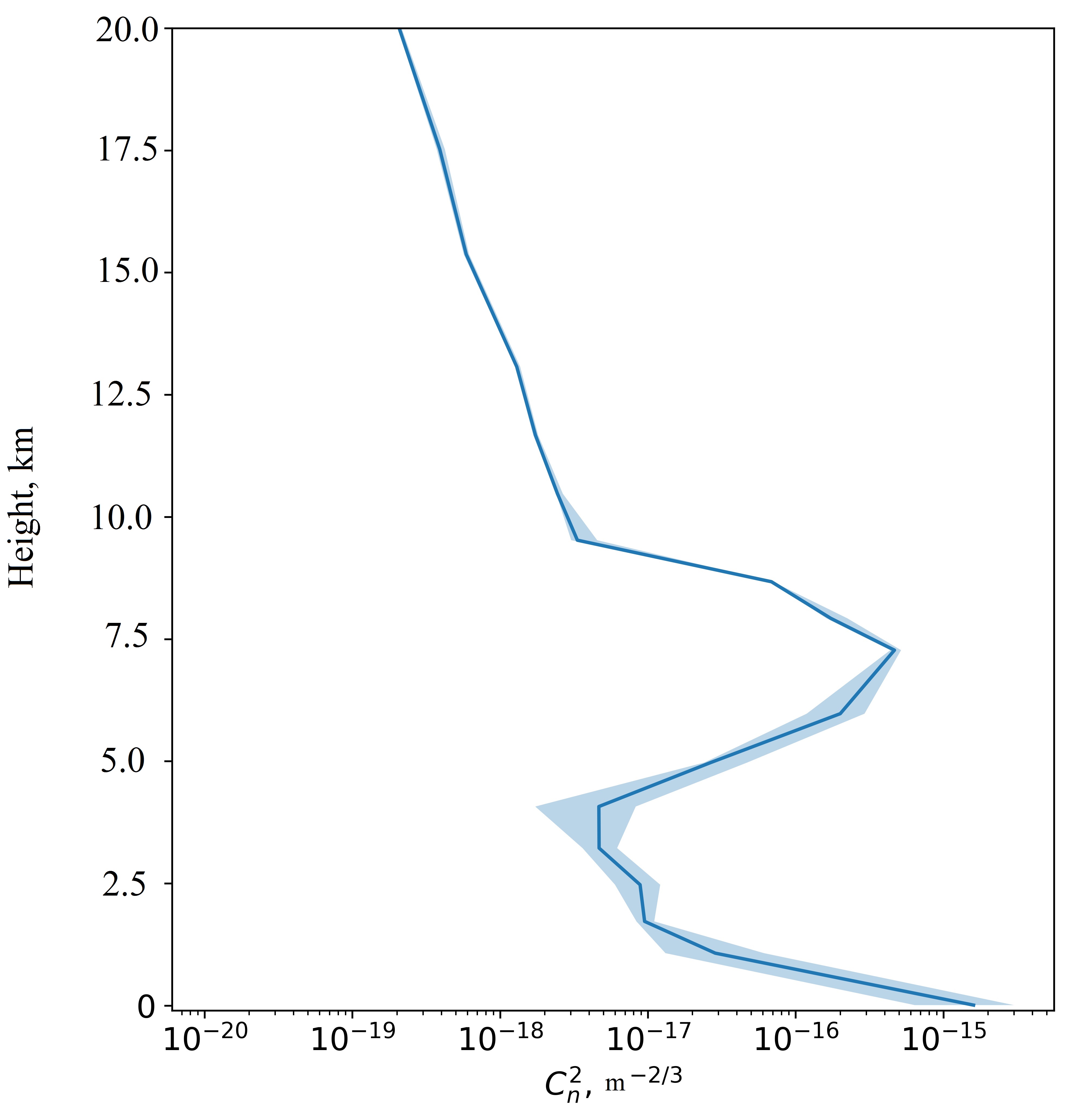}} (\textbf{a}) $a$=1.64, $b$=42, $r_0$ = 5.66 cm, $\beta$ = 1.79 arc sec \\
\end{minipage}
\vfill
\begin{minipage}[h]{0.6185\linewidth}
\center{\includegraphics[width=1\linewidth]{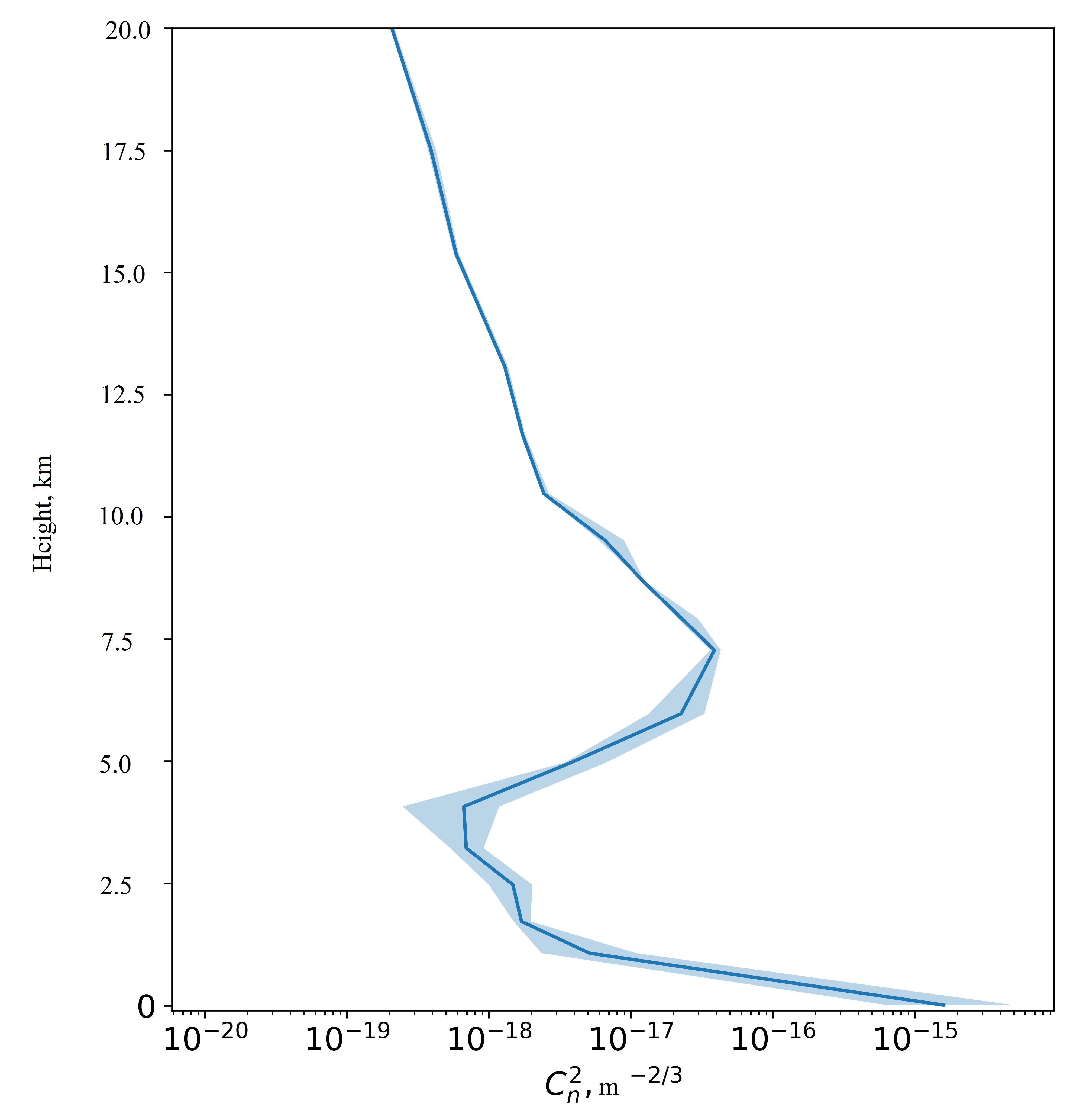}} (\textbf{b}) $a_{new}$=0.95, $b_{new}$=10, $r_0$ = 8.20 cm, $\beta$ = 1.23 arc sec\\
\end{minipage}
\vfill
\caption{Calculated vertical profile of optical turbulence at the Peak Terskol Observatory, 18 - 19 March 2023. Shading corresponds to the interval between the first and the third quartiles estimated for all nights in March 2023}
\label{2dfgg1}
\end{figure}

Use of standard coefficients $a$=1.64, $b$=42, $c$=0.506 and $d$=50 gives mediocre results. In particular, we calculated a vertical profile $C_n^2(z)$ for 18 - 19 March, 2023 (figure \ref{2dfgg1} a)). By integrating this profile using formula \ref{edqS}, we obtained the average estimation of $\beta$ parameter:

\begin{equation}\label{edqS}
\beta_m =\frac{0.98 \lambda}{\left(0.423 sec\alpha_s \left(\frac{2 \pi}{\lambda}\right)^2 \int_0^H C_n^2(z)dz \right)^{-3/5}},
\end{equation}

where $H$ is the height of the optically active atmosphere, $\alpha_s$ is the zenith angle, $\lambda$ is the wavelength of light. 
In average, the model value of $\beta_m$ is higher than the measured $\beta$ and and the estimated $C_n^2$ in the upper atmospheric layers are higher than expected values. In order to correct the vertical profile of $ C_n^2(z)$ we calculated new parameterization coefficients $a_{new}$ and $b_{new}$ by finding the best fitting model linking measured and model near-surface values of $C_n^2$.

Specifically, new coefficients were calculated by minimizing the standard deviation between the model values $C_n^2$ and the measured, sonic, values of  $C_n^2 (m)$:
\begin{equation}\label{ffequS}
\sum_i ^{T_L}(C_{ni}^2 (m)  - C_{ni}^2)^2 \rightarrow min,
\end{equation}
where $C_{ni}^2 (m) $ is the measured value of $C_n^2$ in the surface layer, $C_{ni}^2$ is the model (calculated) value of $C_n^2(surf)$. Summation is performed over the entire time implementation, the length of which is $T_L$. Using this minimization condition, the weight coefficients $a_{new}$ and $b_{new}$ are estimated. New parametrization coefficients at the Terskol Peak Observatory differ significantly from standard values: $a_{new}$=0.95, $b_{new}$=10. Figure \ref{2dfgg1} b) shows corrected vertical profile of $C_n^2$ with the new parametrization coefficients. This profile corresponds to 19 March, 2023.  

\begin{figure}
\centering
\begin{minipage}[h]{0.655\linewidth}
\includegraphics[width=1\linewidth]{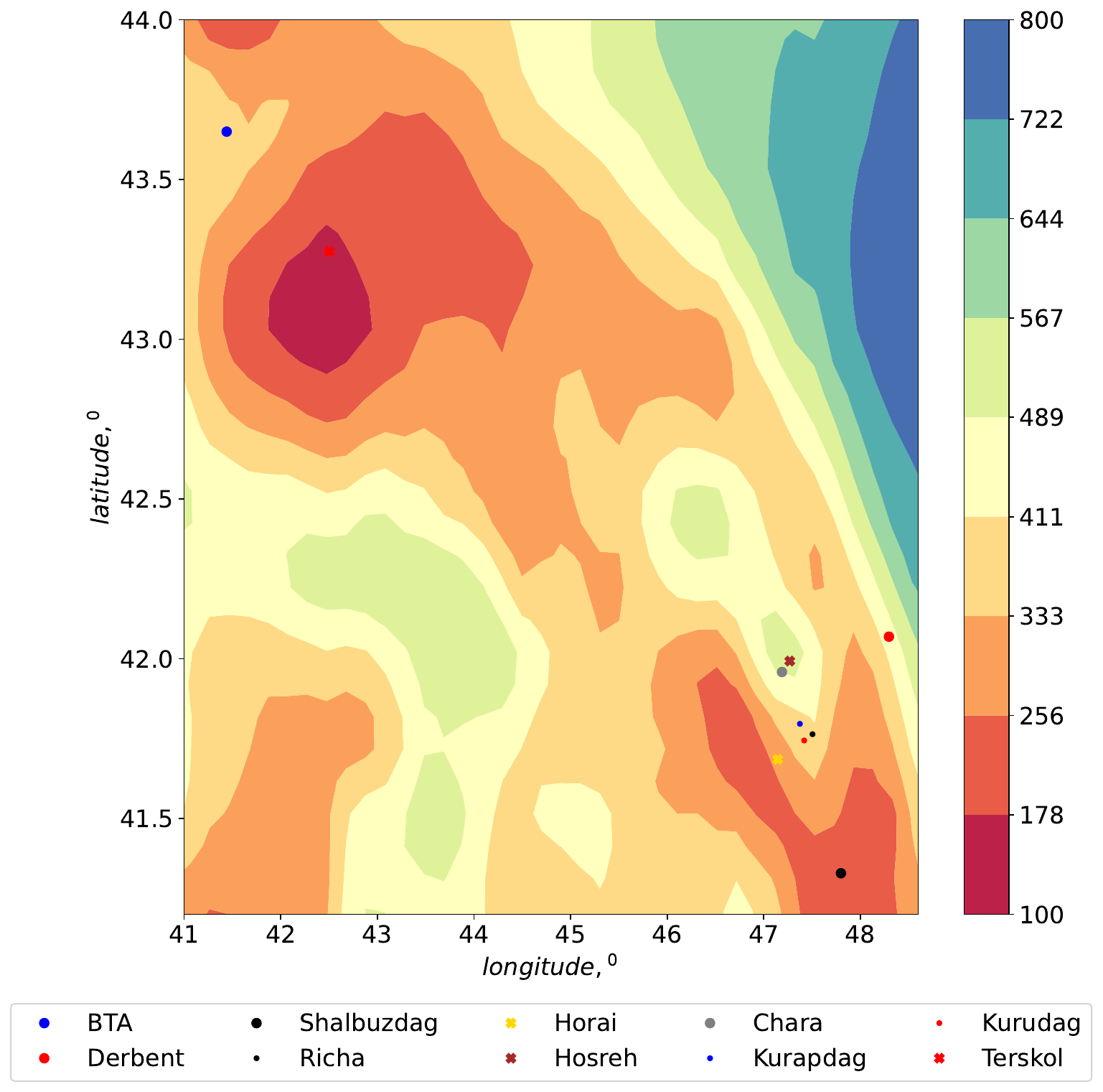} (\textbf{a}) Average atmospheric conditions, 2013 - 2023 \\
\end{minipage}
\vfill
\begin{minipage}[h]{0.655\linewidth}
\includegraphics[width=1\linewidth]{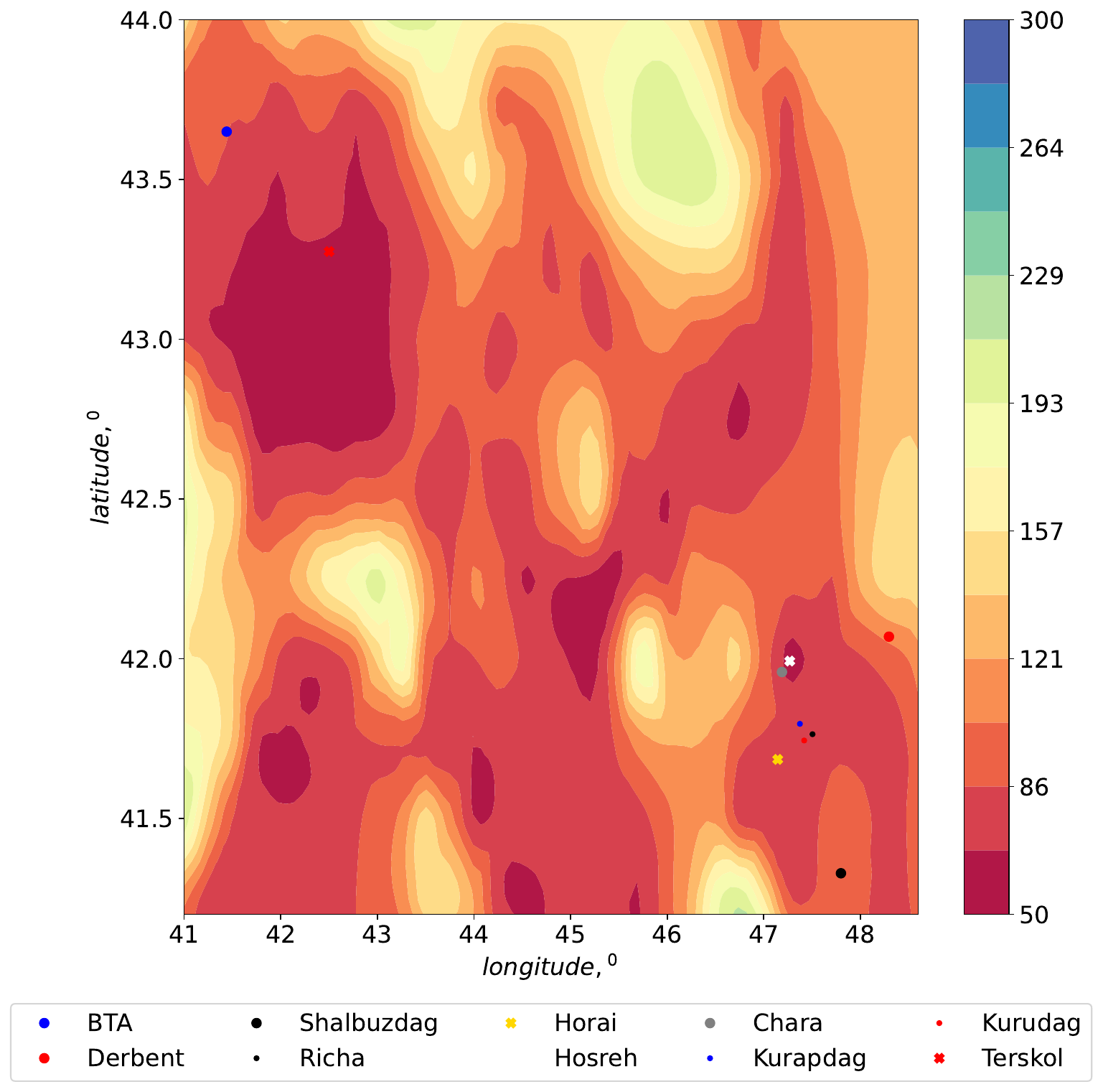} (\textbf{b})  19 March, 2023\\
\end{minipage}
\vfill
\caption{Distributions of atmospheric boundary layer height (in meters) within the Peak Terskol Observatory region.}
\label{2dfedre1}
\end{figure}

One of the most important characteristics of optical turbulence is the height of atmospheric boundary layer (ABL). For determination of the ABL height we have used a well-known method based on the profile of the bulk Richardson number. The ABL height is defined as the height where the bulk Richardson number exceeds a critical threshold. From figure \ref{2dfgg1}, we can see that the new coefficients used to parameterize the night variations of $C_n^2$ made it possible to reduce the height of ABL.  Obtained night-time heights of the atmospheric boundary layer ranges from 100 to 300 m. As a result, the estimated strength of optical turbulent fluctuations decreases more rapidly with height at night above Terskol Peak. We associate this decrease of turbulent fluctuations strength within the lower atmospheric layer above the Terskol Peak with the low heights of the atmospheric boundary layer at night. Considering a large region, the minimum height of the atmospheric boundary layer is observed above Terskol peak not only on individual nights, but also on average, over long time period (figure \ref{2dfedre1}). The high quality of astronomical images at the Terskol Peak is related to the low height of the atmospheric boundary layer. On some nights, its height is limited by 50 - 100 m. Mt Kurapdag and Big Telescope Alt-Azimuthal are located in worse atmospheric conditions, the mean night-time heights of the atmospheric boundary layer are $\sim$ 250 - 300 m and 250 - 350 m respectively (figure \ref{2dfedre1} a)). We believe that observed character of vertical changes is in good agreement with theoretical and experimental data on the evolution of the atmospheric boundary layer.
 
\begin{figure}
\centering
\begin{minipage}[h]{0.65\linewidth}
\center{\includegraphics[width=1\linewidth]{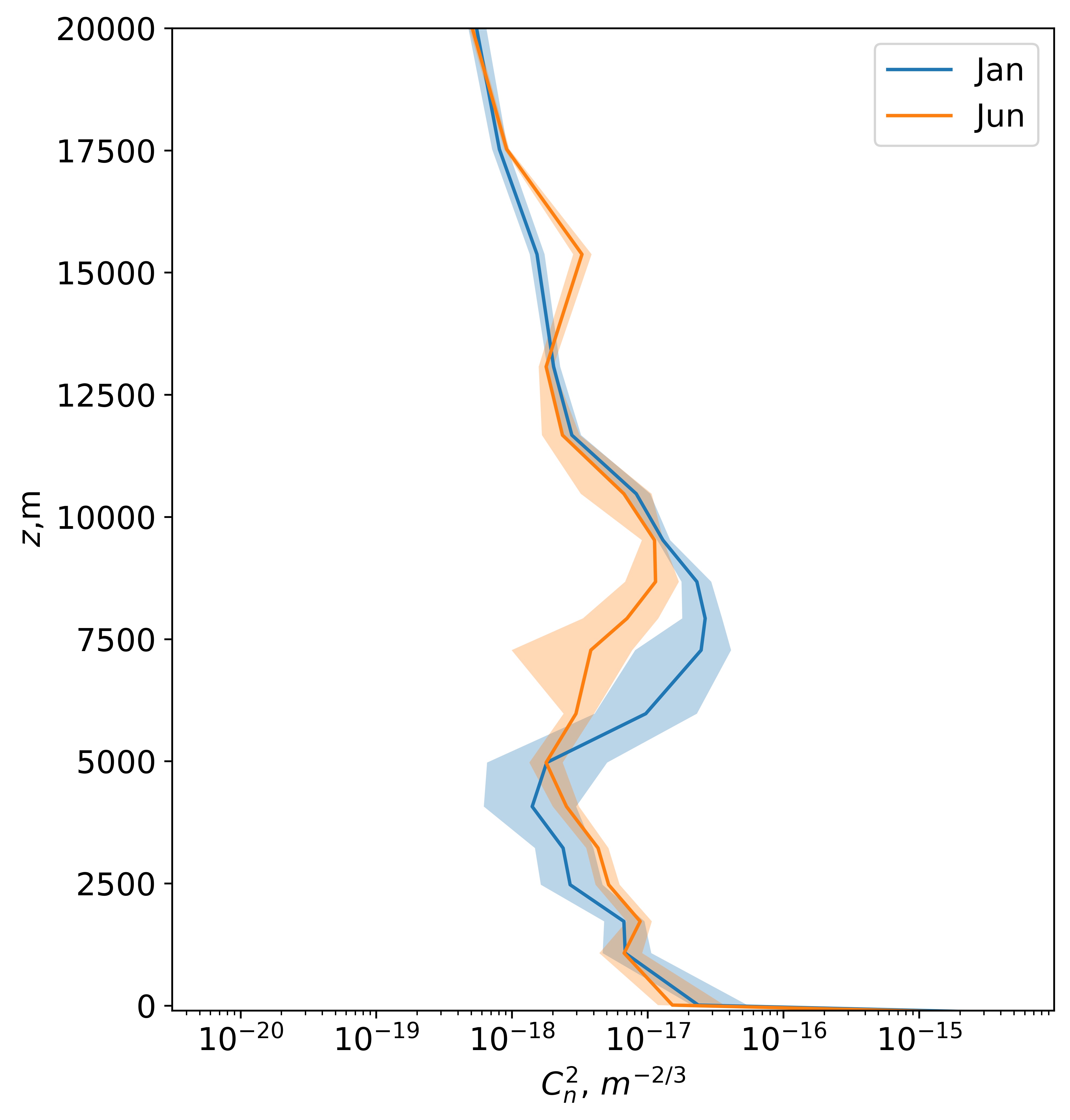}} (\textbf{a}) Peak Terskol Observatory,  2023. Blue line corresponds to January. Orange line corresponds to June.  Shading corresponds to the interval between the first and the third quartiles \\
\end{minipage}
\vfill
\begin{minipage}[h]{0.65\linewidth}
\center{\includegraphics[width=1\linewidth]{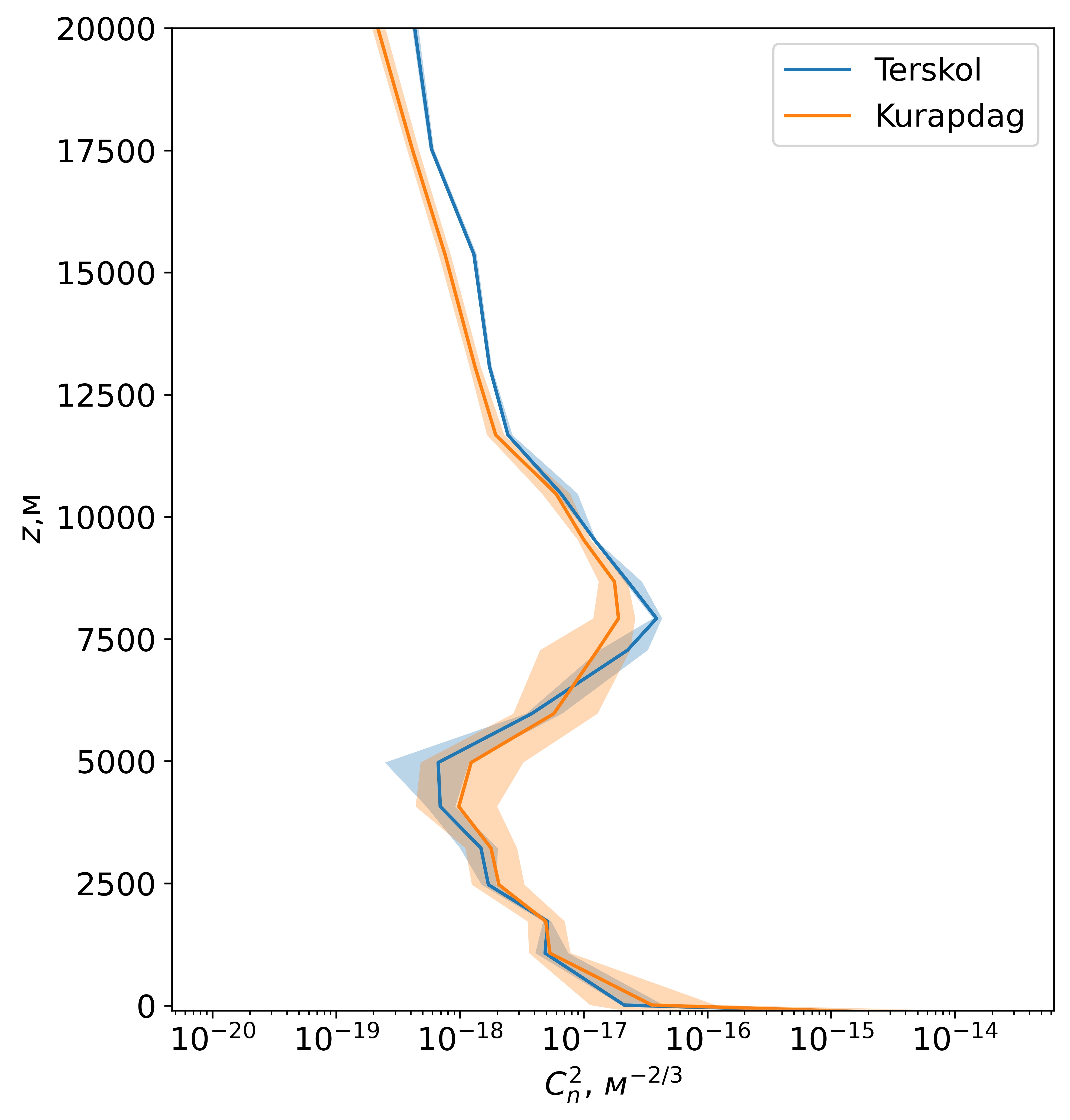}} (\textbf{b}) Mt. Kurapdag, 2013 - 2023. Blue line corresponds to the Peak Terskol Observatory. Orange line corresponds to Mt. Kurapdag.  \\
\end{minipage}
\vfill
\caption{Night-time vertical profiles of optical turbulence above the Peak Terskol Observatory and Mt.Kurapdag}
\label{2dfd1}
\end{figure}

The data of optical and meteorological measurements as well as the fields of wind speed and air temperature extracted from the re-analysis Era-5 , served as the basis for calculation of the average profiles of optical turbulence above the observatory and Mt.Kurapdag. In figure \ref{2dfd1} a) we demonstrate calculated vertical profiles of optical turbulence above the Peak Terskol Observatory averaged over January, 2023 and June, 2023. 

Analyzing these profiles, it can be noted that the structure of turbulence in the upper atmospheric layers changes significantly in summer. This fact is indicated by deformations of vertical profiles of optical turbulence. On the one hand, these deformations manifest in a greater vertical uniformity of optical turbulence and suppression of turbulent fluctuations in the surface layer of the atmosphere. On the other hand, the turbulence strength in the upper atmospheric layers decreases in summer.

Also, using Era-5 reanalysis data and identified patterns in optical turbulence behaviour at the Peak Terskol Observatory site we calculated averaged vertical profiles of $C_n^2(z)$ above the Mt. Kurapdag. For comparison, the figure \ref{2dfd1} b) shows vertical profiles above Terskol Peak and Kurapdag Mountain over 2013 - 2023. Analysis of these vertical profiles indicates the presence of an intense lower turbulence layer. It should be noted that within the surface layer, the intensity of turbulence, in terms of $C_n^2$, at the Mt. Kurapdag site exceeds by more than an order of magnitude the values of $C_n^2$ observed at the Peak Terskol Observatory. Moreover, at the Mt. Kurapdag site, turbulence in this atmospheric layer is characterized by high intensity regardless of the parameterization coefficients. The nature of the turbulence is largely due to strong vertical wind speed shears near the ground.

\section{Estimation of the Fried parameter at the sites with good astroclimatic conditions}

\begin{figure}
\includegraphics[scale=1]{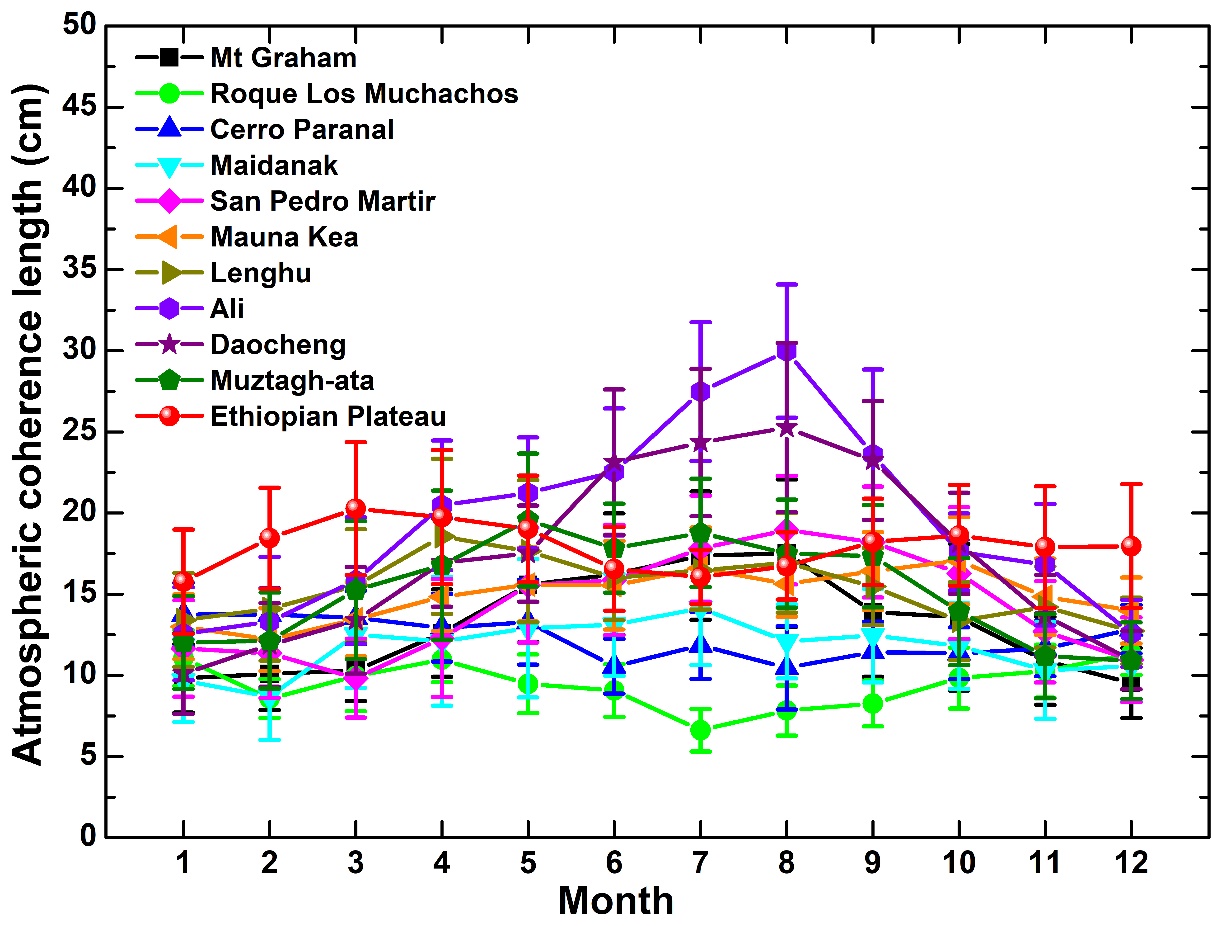}
\caption{Variations of the Fried parameter $r_0$=$0.98 \lambda /\beta$}
 \label{Fried}
\end{figure}

Figure \ref{Fried} displays the monthly averaged Fried parameter estimated using Era-5 reanalysis data at the ten astronomical sites. European Southern Observatory and Tibetan Plateau and the Ethiopian Plateau are some of the most good sites. In contrast to the Fried parameter of the  European Southern Observatory and Tibetan Plateau (TP) sites, the Ethiopian Plateau has a different pattern, possibly owing to different climate patterns. During the year, the Ethiopian Plateau atmospheric coherence length was high in winter and low in summer. In winter, atmospheric coherence length at the Ethiopian Plateau is better than that at the ESO and TP sites, indicating that the Ethiopian Plateau has high potential conditions that are comparable to some of the best astronomical observatories in the world. Estimated Fried parameter at the Ali site located in western China and the Tibetan plateau, one of the best astronomical sites, is about 20 cm in the best months. These  sites are presently under studies for astronomy \cite{Hickson2020}. The Era-5 derived values of the Fried parameter corresponding to spatial coherence length are consistent with measured values at the Maidanak Observatory \citep{Tillayev2023}.

\begin{figure}
\includegraphics[scale=1.425]{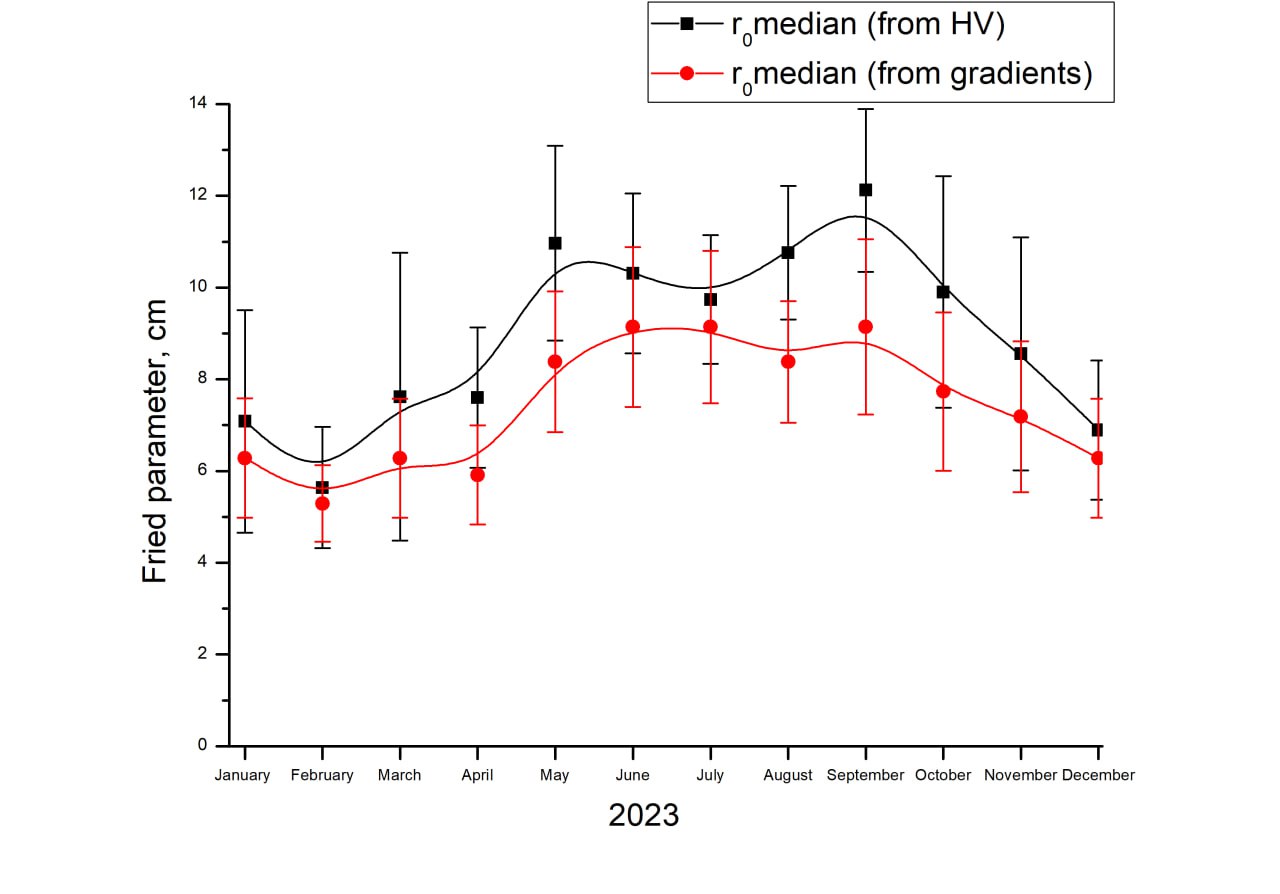}
\caption{Variations of the Fried parameter $r_0$ at the Terskol Peak Observatory}
 \label{Friedf}
\end{figure}

For comparison, we also compare the variation of the Fried parameter at  the site of Peak Terskol. Figure \ref{Friedf} shows variations of the Fried parameter $r_0$ at the Terskol Peak Observatory. As you can see from the figure, the site of Terskol Peak has a high image quality close to the best places in the world. In particular, in summer $\beta$ is about 1.0 - 1.2  arc sec, in terms of $r_0$, these values correspond to 8.3 - 10.0 cm.  In addition to the average monthly values of the Fried parameter obtained using the gradient method, the figure shows the results of applying the Hufnagel–Vally (HV) climate model to describe vertical changes in $C_n^2$ \citep{HV2023}. 
The HV model is based on the processing of experimental data and, on average, describes the structure of optical turbulence well. Nevertheless, detailed information about the vertical distribution of optical turbulence cannot be reflected using this model. Analysis of the figure shows that the gradient model and the model give similar results in terms of $r_0$. 

\section {Discussion}


In this paper, we present the results of studies of the optical turbulence in the atmospheric boundary layer and free atmosphere, up to a height of 20 km. In order to determine the structural constant of turbulent fluctuations of the air refractive index, we refined the parameterization scheme for calculation of this quantity. In the region of interest, this turbulent parameter was calculated based on a comparison of reanalysis data, mast meteorological measurements and observations of image scintillation. 
In particular, the processing of mast measurement data made it possible not only to estimate $C_n^2$ statistics in the atmospheric surface layer, but also to select optimal parameterization coefficients that determine the contribution of wind speed shear instability to the formation of turbulence. Using these coefficients,  we estimated  $C_n^2$ values at different heights in the atmospheric boundary layer more accurately.  This made it possible to refine the shapes of the vertical profiles of optical turbulence. The correctness of our approach based on the using constant parametrization coefficients (within only atmospheric boundary layer)  is supported by the fact that the structure of turbulence in the atmospheric boundary layer should be determined by processes in the surface layer of the atmosphere, at least for long time periods.

Correctness of the calculations is also confirmed by the results of estimating the height of the atmospheric boundary layer. Figures \ref{2dfedre1} show distributions of the atmospheric boundary layer height (in meters) within the Peak Terskol Observatory region. Analysis of these distributions allows us to notice that a fairly thin boundary layer is formed above the Observatory. This suggests that turbulence develops predominantly in the layer near the ground. During the measurements, the thickness of the atmospheric boundary layer was minimal (compared to average atmospheric conditions (\ref{2dfedre1} a)). 


It can be noted that the gradient method outputs are in good agreement with the data of optical measurements of turbulence at different height levels in the atmosphere, at least for the Terskol Peak Observatory. Based on the selected profile calculation scheme for the Terskol Peak Observatory, we estimated the optical turbulence characteristics for other sites as well.  Analysis of spatial distribution of the atmospheric boundary layer height, as well as calculated profiles of optical turbulence, can serve as the basis for choosing optimal site for a ground-based astronomical telescope. Taking into account the recent data on the spatial distribution of astroclimatic characteristics within this macroregion, the following areas suitable for optical telescopes are located: in mountain Dagestan, including Gunib, as well as Shalbuzdag and Peak Terskol Observatory. The location of Mt. Kurapdag is not good due to strong surface winds and strong surface optical turbulence, which is the main reason of high values of $\beta$ (2- 3 $''$).



\section{Conclusions}


In this work we demonstrated that, using the gradient method and additional measurement data, we can estimate the vertical profiles of optical turbulence more accurately. The increase of accuracy of optical turbulence profiles within the lower atmosphere is due to using both measured values of $C_n^2$ and applying new parametrization coefficients between vertical shears of wind speed and model values of $C_n^2$. Moreover, the gradient method takes into account variations in the height of the atmospheric boundary layer estimated from re-analysis data. Estimations of the atmospheric boundary layer height and the vertical profiles of wind shears are important for obtaining smooth vertical profiles of $C_n^2$ and $L_0$. Above the atmospheric boundary layer, vertical profiles of optical turbulence are reconstructed only on the basis of vertical profiles of wind speed shears with parameterization coefficients $a$=1.64 and $b$=42, that describe $C_n^2$ variations well for weak turbulence. Thus, we can conclude that the hypotheses of optical turbulence  embedded in the gradient method are valid as, in average,  used atmospheric models converge to the measurements: the measured and model values of $\beta$ are practically identical.  

For the first time, by linking simulation outputs to measurement data, we have obtained representative optical turbulence vertical profiles for the sites of Peak Terskol Observatory and the Mt. Kurapdag. We have shown that the Kurapdag mountain should not be considered as a potential site for a new optical telescope.


\begin{thebibliography}{999}
\bibitem[Coulman(1985)]{Coulman1985}
Coulman, C.E. Fundamental and applied aspects of astronomical “seeing”. {\em Annu. Rev. Astron. Astrophys.} {\bf 1985}, {\em 23}, 19-57. Doi:10.1146/annurev.aa.23.090185.000315.

\bibitem[Tokovinin(2023)]{Tokovinin2023}
Tokovinin, A. The Elusive Nature of “Seeing”. {\em 	Atmosphere} {\bf 2023}, {\em 14}, {\em 11}, 1694. Doi:10.3390/atmos14111694.


\bibitem[Ran(2024)]{Ran2024}
Ran, X.; Zhang, L.; Rao, C. The AC-SLODAR: measuring daytime normalized optical turbulence intensity distribution based on slope autocorrelation. {\em Mnras} {\bf 2024}, {\em 528}, {\em 3}, 3981-3991. Doi:10.1093/mnras/stae202.

\bibitem[Hickson (2019)]{Hickson2019}
Hickson, P.; Ma, B.; Shang, Z.;  Xue, S. Multistar turbulence monitor: a new technique to measure optical turbulence profiles. {\em Mnras} {\bf 2019}, {\em 485}, {\em 2}, 2532 – 2545, doi:10.1093/mnras/stz568.

\bibitem[Hickson (2020)]{Hickson2020}
Hickson, P.; Feng, L.; Hellemeier, J. A.; Shen, Z.; Xue, S.; Yao, Y.; Ma, B.; Chen, H.; Yang, R. Optical turbulence at Ali, China – results from the first year of lunar scintillometer observation. {\em Mnras} {\bf 2020}, {\em 494}, {\em 4}, 5992 – 6000, doi:10.1093/mnras/staa1101.

\bibitem[Wilson R.(2002)]{Wilson}
Wilson, R. SLODAR: Measuring optical turbulence altitude with a Shack–Hartmann wavefront sensor. {\em Mon. Not. R. Astron. Soc.} {\bf 2002}, {\em 337}, 103--108.

\bibitem[Deng(2023)]{Deng2023}
Deng, J.; Song, T.-F.; Liu, Y.
A Review of Daytime Atmospheric Optical Turbulence Profile Detection Technology. {\em Chinese Astronomy and Astrophysics} {\bf 2023}, {\em 47}, {\em 2}, 257--284. Doi:10.1016/j.chinastron.2023.06.001.


\bibitem[Wang Z.(2018)]{Wang2018}
Wang, Z.; Zhang, L.; Kong, L.; Bao, H.; Guo, Y.; Rao, X.; Zhong, L.; Zhu, L.; Rao, C. A modified S-DIMM+: Applying additional height grids for characterizing daytime seeing profiles. {\em Mon. Not. R. Astron. Soc.} {\bf 2018}, {\em 478}, 1459--1467.

\bibitem[Subramanian(2023)]{Subramanian2023}
Subramanian, S.K.; Rengaswamy, S. Forward modelling of turbulence strength profile estimation using S-DIMM+. {\em 	
Proceedings of SPIE} {\bf 2023}, {\em 12638}, 1263812. Doi:10.1117/12.2671735.


\bibitem[Shikhovtsev(2022)]{Shikhovtsev2022}
Shikhovtsev, A.Yu. A Method of Determining Optical Turbulence Characteristics by the Line of Sight of an Astronomical Telescope. {\em Atmospheric and Oceanic Optics} {\bf 2022}, {\em 35}, {\em 3}, 303-309. Doi:10.1134/S1024856022030149.


\bibitem[Joo(2024)]{Joo2024}
Joo, J.Y.; Ha, H.S.; Lee, J.H.; Kim, Y.S.; Butterley, T. SLODAR System Development for Vertical Atmospheric Disturbance Profiling at Geochang Observatory. {\em Current Optics and Photonics} {\bf 2024}, {\em 8}, {\em 1}, 30-37. Doi:10.3807/COPP.2024.8.1.30.


\bibitem[Bennoui(2023)]{Bennoui2023}
Bennoui, F.; Bahloul, D. Estimation of the Atmospheric Turbulence Parameters Using the Angle-of-Arrival Covariance Function. {\em Atmospheric and Oceanic Optics} {\bf 2023}, {\em 36}, {\em 5}, 569-577. Doi:10.1134/S1024856023050202.


\bibitem[Griffiths(2024)]{Griffiths2024}
Griffiths, R.; Bardou, L.; Butterley, T.; Osborn, J.; Wilson, R.; Bustos, E.; Tokovinin, A.; Le Louarn, M.; Otarola, A. A comparison of next-generation turbulence profiling instruments at
Paranal. {\em Monthly Notices of the Royal Astronomical Society} {\bf 2024}, {\em 529}, {\em 1}, 320–330. Doi:10.1093/mnras/stae434.


\bibitem[Zhang(2022)]{Zhang2022}
Zhang, K.; Luo, T.; Wang, F.-F.; Sun, G.; Liu, Q.; Qing, C.; Li, X.-B.; Weng, N.-Q.; Zhu, W.-Y. Influence of low clouds on atmospheric refractive index structure constant based on radiosonde data. {\em 	
Acta Physica Sinica} {\bf 2022}, {\em 71}, 089202. Doi:10.7498/aps.71.20211792.


\bibitem[Bolbasova(2019)]{Bol2019}
Bolbasova, L.A.; Shikhovtsev, A.Yu.; Kopylov, E.A.; Selin, A.A.; Lukin, V.P.; Kovadlo, P.G. Daytime optical turbulence and wind speed distributions at the Baikal Astrophysical Observatory. {\em  Monthly Notices of the Royal Astronomical Society} {\bf 2019}, {\em 482}, 2619–2626. Doi:10.1093/mnras/sty2706.


\bibitem[Qing(2020)]{Qing2020}
Qing, C.; Wu, X.; Li, X.; Luo, T.; Su, C.; Zhu, W. Mesoscale optical turbulence simulations above Tibetan Plateau: first attempt. {\em Optics Express} {\bf 2020}, {\em 28}, {\em 4}, 4571--4586, doi:10.1364/OE.386078.

\bibitem[Bi (2023)]{Bi2023}
Bi, C.; Qing, C.; Qian, X.; Luo, T.; Zhu, W.; Weng, N. Investigation of the Global Spatio-Temporal Characteristics of Astronomical Seeing. {\em Remote Sens.} {\bf 2023}, {\em 15}, 2225, doi:10.3390/rs15092225.

\bibitem[Yang(2021)]{Yang2021}
Yang, Q.; Wu, X.; Han, Y.; Chun, Q.; Wu, S.; Su, C.; Wu, P.; Luo, T.; Zhang, S. Estimating the astronomical seeing above Dome A using Polar WRF based on the Tatarskii equation. {\em Opt. Express} {\bf 2021}, {\em 29}, 44000--44011. http://doi.org/10.1364/OE.439819.

\bibitem[Quatresooz(2023)]{Quatresooz2023}
Quatresooz, F.; Griffiths, R.; Bardou, L.; Wilson, R.; Osborn, J.; Vanhoenacker-Janvier, D.; Claude, O. Continuous daytime and nighttime forecast of atmospheric optical turbulence from numerical weather prediction models. {\em Optics Express} {\bf 2023}, {\em 31}, {\em 21}, 33850--33872. Doi:10.1364/OE.500090.

\bibitem[Wu(2024)]{Wu2024}
Wu, X.-Q.; Xiao, C.-Y.; Esamdin, A.; Xu, J.; Wang, Z.-W.; Xiao, L. Quantitative Analysis of Seeing with Height and Time at Muztagh-Ata Site Based on ERA5 Database. {\em Research in Astronomy and Astrophysics} {\bf 2024}, {\em 24}, {\em 1}, 015006. Doi:10.1088/1674-4527/ad057d.


\bibitem[Cuevas(2024)]{Cuevas2024}
Cuevas, O.; Marin, J.C.; Blazquez, J.; Meyer, C. Combining $C_n^2$ models to forecast the optical turbulence at Paranal. {\em Monthly Notices of the Royal Astronomical Society} {\bf 2024}, {\em 529}, {\em 3}, 2208 - 2219. Doi:10.1093/mnras/stae630.


\bibitem[Bi (2024)]{Bi2024}
Bi, C.; Qing, C.; Qian, X.; Zhu, W.; Luo, T.; Li, X.; Cui, S.; Weng, N. Astroclimatic parameters characterization at lenghu site with ERA5 products. {\em Monthly Notices of the Royal Astronomical Society} {\bf 2024}, {\em 527}, {\em 3}, 4616 - 4631, doi:10.1093/mnras/stad3414.



\bibitem[Macatangay (2024)]{Macatangay2024}
Macatangay, R.;  Rattanasoon, S.; Butterley, T.; Bran, S. H.; Sonkaew, T.; Sukaum, B.; Sookjai, D.; Panya, M.; Supasri, T. Seeing and turbulence profile simulations over complex terrain at the Thai National Observatory using a chemistry-coupled regional forecasting model. {\em Monthly Notices of the Royal Astronomical Society} {\bf 2024}, {\em 530}, {\em 2}, 1414 - 1423, doi:10.1093/mnras/stae727.


\bibitem[Masciadri (2001)]{Masciadri2001A}
Masciadri, E.; Jabouille, P. Improvements in the optical turbulence parameterization for 3D simulations in a region around a telescope. {\em Astronomy and Astrophysics} {\bf 2001}, {\em 376}, {\em 2}, 727 – 734, doi:10.1051/0004-6361:20010999.


\bibitem[Turchi (2017)]{Turchi2017}
Turchi, A.; Masciadri, E.; Fini, L. Forecasting surface-layer atmospheric parameters at the Large Binocular Telescope site. {\em Monthly Notices of the Royal Astronomical Society} {\bf 2017}, {\em 466}, {\em 2}, 1925 - 1943, doi:10.1093/mnras/stw2863.

\bibitem[Shikhovtsev (2024)]{Shikhovtsev2024PASJ}
Shikhovtsev, A. Reference optical turbulence characteristics at the Large Solar Vacuum Telescope site. {\em Publications of the Astronomical Society of Japan} {\bf 2024}, psae031, doi:10.1093/pasj/psae031.

\bibitem[Bolbasova(2023)]{Bolbasova2023}
Bolbasova, L.A.; Kopylov E.A. Long-Term Trends of Astroclimatic Parameters above the Terskol Observatory. {\em Atmosphere} {\bf 2023}, {\em 14}, {\em 8}, 1264. Doi:10.3390/atmos14081264.


\bibitem[Khaikin(2022)]{KhaikinESMT}
Khaikin V.B.; Shikhovtsev A.Yu.; Shmagin V.E.; Lebedev M.K.; Kopylov E.A.; Lukin V.P.; Kovadlo P.G. Eurasian Submillimeter Telescopes (ESMT) project. Possibility of submm image quality improvement using adaptive optics. {\em Zhurnal radioelektroniki [Journal of Radio Electronics]} {\bf 2022}, {\em 7}, Doi:10.30898/1684-1719.2022.7.9.


\bibitem[Shikhovtsev(2024)]{Shikhovtsev2024}
Shikhovtsev, A.Yu.; Kovadlo P.G. Statistical estimations of the vapor content and optical thickness of the atmosphere using reanalysis and radiosonding data as applied to millimeter telescopes. {\em Optika Atmosfery i Okeana} {\bf 2024}, {\em 37}, {\em 2}, 169-175. Doi:10.15372/AOO20240212.

\bibitem[Hersbach(2020)]{Hersbach2020}
Hersbach, H.; Bell, B.; Berrisford, P.; Hirahara, S.; Horányi,A.; Muñoz-Sabater, J.; Nicolas,J.; Peubey, C.; Radu, R.; Schepers, D.; Simmons, A.; Soci, C.; Abdalla, S.; Abellan, X.; Balsamo, G.; Bechtold, P.; Biavati, G.; Bidlot, J.; Bonavita, M.; De Chiara, G.; Dahlgren, P.; Dee, D.; Diamantakis, M.; Dragani, R.; Flemming, J.; Forbes, R.; Fuentes, M.; Geer, A.; Haimberger, L.; Healy, S.; Hogan, R.J.; Hólm, E.; Janisková, M.; Keeley, S.; Laloyaux,P.; Lopez,P.; Lupu, C.; Radnoti, G.; de Rosnay, P.; Rozum, I.; Vamborg, F.; Villaume, S.; Thépaut, J.-N. The ERA5 global reanalysis. {\em  Quarterly Journal of the Royal Meteorological Society} {\bf 2020}, {\em 146}, {\em 730}, 1999-2049, doi: 10.1002/qj.3803.


\bibitem[Huang(2021)]{Huang2021}
Huang, J.; Yin, J.; Wang, M.; He, Q.; Guo, J.; Zhang, J.; Liang, X.; Xie, Y. Evaluation of Five Reanalysis Products With Radiosonde Observations Over the Central Taklimakan Desert During Summer. {\em Earth and Space Science} {\bf 2021}, {\em 8},  {\em 5}, 2021EA001707. Doi:10.1029/2021EA001707.


\bibitem[Rao (2024)]{Rao2024}
Rao, P.; Wang, F.; Yuah, X.; Liu, Y.; Jiao, Y. Evaluation and comparison of 11 sets of gridded precipitation products over the Qinghai-Tibet Plateau. {\em Atmospheric Research} {\bf 2024}, {\em 302}, 107315, doi:10.1016/j.atmosres.2024.107315.

\bibitem[Townson (2015)]{Townson2015}
Townson, M.J; Kellerer A.; Saunter C.D. Improved shift estimates on extended Shack–Hartmann wavefront sensor images. {\em Monthly Notices of the Royal Astronomical Society} {\bf 2015}, {\em 452}, 4022–4028, doi:10.1093/mnras/stv1503.

\bibitem[Potanin(2022)]{Potanin2022}
Potanin, S.A.; Kornilov, M.-V.; Savvin, A.D.; Safonov, B.S.; Ibragimov, M.A.; Kopylov, E.A.;  Nalivkin, M.A.; Shmagin, V.E.; Huy, L.X.; Thao, N.T. A Facility for the Study of Atmospheric Parameters Based on the Shack–Hartmann Sensor. {\em Astrophysical Bulletin} {\bf 2022}, {\em 77}, 214–221. Doi:10.1134/S1990341322020067.


\bibitem[Kornilov (2014)]{Kornilov2014}
Kornilov V.; Safonov B.; Kornilov M.; Shatsky N.; Voziakova O.; Potanin S.; Gorbunov I.; Senik V.; Cheryasov D. Study on atmospheric optical turbulence above mount Shatdzhatmaz in 2007-2013. {\em Publications of the Astronomical Society of the Pacific} {\bf 2014}, {\em 126}, {\em 939}, 482-495, doi:10.1086/676648.

\bibitem[Panchuk (2011)]{Panchuk2011}
Panchuk V.E.; Afanas'ev V.L. Astroclimate of Northern Caucasus-myths and reality. {\em Astrophysical Bulletin} {\bf 2011}, {\em 66}, {\em 2}, 233-254, doi:10.1134/S199034131102009X.


\bibitem[Odintsov(2019)]{Odintsov Atmosphere}
Odintsov, S.L.; Gladkikh, V.A.; Kamardin, A.P.; Nevzorova, I.V. Determination of the Structural Characteristic of the Refractive Index of Optical Waves in the Atmospheric Boundary Layer with Remote Acoustic Sounding Facilities. {\em  Atmosphere} {\bf 2019}, {\em 10}, 711. doi:10.3390/atmos10110711.

\bibitem[Lukin(2015)]{Lukin2015}
Lukin, V.P.; Botygina, N.N.; Gladkikh, V.A.; Emaleev, O.N.; Konyaev, P.A.; Odintsov, S.L.; Torgaev, A.V. Joint measurements of atmospheric turbulence level with optical and acoustic meters.  {\em Atmospheric and Oceanic Optics} {\bf 2015}, {\em 28}, {\em 3}, 254 - 257. Doi: 10.1134/S1024856015030100.


\bibitem[Rao (2023)]{Rao2023}
Rao, R. Effect of Outer Scale of Atmospheric Turbulence on Imaging Resolution of Large Telescopes. {\em Guangxue Xuebao/Acta Optica Sinica} {\bf 2023}, {\em 43}, {\em 24}, 2400001, doi:10.3788/AOS231259.



\bibitem[Tillayev (2023)]{Tillayev2023}
Tillayev, Y.; Azimov, A.; Ehgamberdiev, S.; Ilyasov, S. Astronomical   Seeing   and  Meteorological  Parameters  at  Maidanak Observatory. {\em Atmosphere} {\bf 2023}, {\em 14}, 199, doi:10.3390/atmos14020199.

\bibitem[Mahmood (2022)]{HV2023}
Mahmood, D.A.; Naif, S.S.; Al-Jiboori, M.H.; Al-Rbayee, T. Improving Hufnagel-Andrews-Phillips model for prediction $C_n^2$ using empirical wind speed profiles. {\em JASTP} {\bf 2022}, {\em 240}, 105952, doi:10.1016/j.jastp.2022.105952.

\end{thebibliography}
\end{document}